\documentclass[twocolumn,twocolappendix,nofootinbib]{openjournal}

\usepackage[utf8]{inputenc}
\usepackage[T1]{fontenc}

\usepackage{amsmath}
\usepackage{natbib}
\usepackage{newtxtext,newtxmath}
\usepackage{graphicx}
\usepackage[dvipsnames]{xcolor}
\makeatletter
\def\prismkernellabel#1{%
  \@bsphack
  \begingroup
    \UseHookWithArguments{label}{1}{#1}%
    \protected@write\@auxout{}{%
      \string\newlabel{#1}{{\@currentlabel}{\thepage}{\@currentlabelname}{\@currentHref}{\@kernel@reserved@label@data}}%
    }%
  \endgroup
  \@esphack
}
\let\label\prismkernellabel
\makeatother
\usepackage[colorlinks,allcolors=blue]{hyperref}

\AtBeginDocument{%
}
\makeatletter
\newif\ifprismlastpagelabelwritten
\prismlastpagelabelwrittenfalse
\def\lastpage@putlabel{%
 \ifprismlastpagelabelwritten\else
  \global\prismlastpagelabelwrittentrue
  \if@filesw
   \begingroup
    \advance\c@page\m@ne
    \immediate\write\@auxout{\string\newlabel{LastPage}{{}{\thepage}{}{}{}}}%
   \endgroup
  \fi
 \fi
}
\makeatother
\DeclareFontFamily{U}{lasy}{}
\DeclareFontShape{U}{lasy}{m}{n}{<5><5.5><6><7><8><9>gen * lasy <10><10.95><12><14.4><17.28><20.74><24.88>lasy10}{}
\DeclareFontShape{U}{lasy}{b}{n}{<->ssub*lasy/m/n}{}

\makeatletter

\newcommand{\Rmnum}[1]{\expandafter\@slowromancap\romannumeral #1@}

\newcommand{\gsim}{\lower0.6ex\vbox{\hbox{$\buildrel{\textstyle >}\over{\sim}\ $}}}

\def\mpch{h^{-1}{\rm Mpc}}

\begin{document}
\journalinfo{The Open Journal of Astrophysics}
\submitted{Version October 5, 2026}

\title[AI-SR Cosmological Simulations V]
{AI-assisted super-resolution cosmological simulations V: Cosmology-aware super-resolution}

\author{
Xiaowen Zhang\textsuperscript{1},
Patrick Lachance\textsuperscript{1},
Rupert A.~C. Croft\textsuperscript{1},
Tiziana Di Matteo\textsuperscript{1},
Francisco Villaescusa-Navarro\textsuperscript{2,3}
}
\affiliation{
\textsuperscript{1} McWilliams Center for Cosmology, Department of Physics, Carnegie Mellon University, Pittsburgh, PA 15213, USA \\
\textsuperscript{2} Center for Computational Astrophysics, Flatiron Institute, New York, NY 10010, USA \\
\textsuperscript{3} Department of Astrophysical Sciences, Princeton University, Princeton, NJ 08544, USA \\
Correspondence: \texttt{xiaowen4@andrew.cmu.edu}
}

\begin{abstract}
We present a cosmology-aware generative model for super-resolution (SR) of cosmological $N$-body simulations in particle phase space. The model jointly generates displacements and velocities, conditioning a generative adversarial network (GAN) based generator and discriminator on evolved low-resolution (LR) fields and five $\Lambda$CDM parameters $(\Omega_m,\Omega_b,h,n_s,\sigma_8)$. We train a single model on 100 matched low-high resolution pairs from the Quijote Latin-hypercube suite at $z = 0$, mapping $512^3$ to $1024^3$ particles in $1\,h^{-1}\,\mathrm{Gpc}$ boxes. On ten cosmologies excluded from SR training, the model reproduces the broad matter distribution and the cosmological variation of clustering and halo abundances. Over $0.1<k<3\,h\,\mathrm{Mpc}^{-1}$, the mean of the per-cosmology maximum absolute power-spectrum errors is $3.6\%$, while FoF halo abundances agree with HR to within $15\%$ in the analyzed mass bins below $10^{14.8}\,h^{-1}M_\odot$. Satellite abundances and occupations retain larger, mass-dependent discrepancies. Corrections measured at one near-fiducial cosmology reduce a shared component of these systematic errors when applied to the other nine test cosmologies. A power spectrum based inference test also shows that calibration reduces the mean shifts in all parameters. These results extend our SR model across cosmologies and provide 6D phase space realizations for subsequent studies of non-linear structure and field-level cosmological inference.
\end{abstract}

\maketitle

\section{Introduction}
\label{sec:intro}

Cosmological $N$-body simulations are a primary tool for forward-modeling the matter and halo distributions measured by galaxy surveys. Stage-IV surveys, including the Dark Energy Spectroscopic Instrument \citep[DESI;][]{DESI2016}, the \textit{Euclid} space telescope \citep{Laureijs2011}, and the Vera C. Rubin Observatory Legacy Survey of Space and Time \citep[LSST;][]{Ivezic2019}, require accurate predictions over large cosmological volumes and into the nonlinear regime. Resolving these small scales in a full $N$-body run requires particle counts that scale as the inverse cube of the resolution element, so producing the thousands to millions of independent realizations needed for covariance estimation and for simulation-based inference \citep{Hahn2023simbig,VillaescusaNavarro2020} is computationally expensive. Modern simulation campaigns address these demands through different choices of volume, resolution, and physical modeling, from the high-resolution hydrodynamic zoom-in simulations of FIRE \citep{Hopkins2014}, to the large-volume hydrodynamic simulation ASTRID \citep{astrid,Zhou2026Astrid}, and the gravity-only AbacusSummit suite designed for large-scale structure studies \citep{Maksimova2021}. GPU-accelerated gravity-only runs such as Farpoint \citep{Frontiere2022}, together with the recent Euclid Flagship simulation \citep{EuclidFlagship2025}, have extended these calculations to trillions of particles. Approximate solvers such as \textsc{COLA} and \textsc{FastPM} reduce the cost of particle evolution through approximations to the gravitational dynamics \citep{Tassev2013,fastpm}, while emulators provide rapid predictions for selected summary statistics \citep[e.g.][]{Nishimichi2019,Angulo2021}. Field and particle outputs are needed for constructing mock catalogs, measuring additional statistics, and applying survey selection effects. These requirements motivate machine learning approaches that enhance the output of inexpensive, low-resolution (LR) runs.

In recent years machine learning has become a versatile tool across cosmology \citep[for reviews see][]{Dvorkin2022,HuertasCompany2023}. Neural networks have been trained to predict non-linear structure directly from linear initial conditions \citep{He2019} and to model halo populations \citep{Berger2019,Bernardini2020}. 
The growth of large simulated training sets such as CAMELS \citep{VillaescusaNavarro2020camels} has further enabled studies of the joint dependence of structure formation on cosmology and
astrophysics. A closely related line of work builds field-level emulators: \citet{KodiRamanah2020} emulate high-resolution (HR) density fields from LR together with HR initial conditions from the Quijote suite, and \citet{Jamieson2023} map the Zel'dovich approximation to evolved $N$-body displacements and velocities across the Quijote Latin-hypercube cosmologies. The latter approach incorporates cosmological dependence through style parameters, with an extension to redshift-dependent evolution presented in \citet{Jamieson2024}.

Restoring small-scale information that is absent from a coarse input, or super-resolution (SR), generally admits multiple HR realizations consistent with the LR field. When the additional small-scale initial modes are unspecified, this motivates learning the conditional distribution of HR realizations given the LR input. Generative adversarial networks \citep[GANs;][]{Goodfellow2014} provide one approach to this problem and have been applied to image SR \citep{Ledig2016,Wang2019}. In cosmology they have been used to synthesize 2D cosmic-web images \citep{Rodriguez2018} and 3D matter density fields \citep{Perraudin2019}, and, more recently, to emulate mass maps and large-scale structure maps across cosmologies \citep{Perraudin2020,Bhambra2025}. Denoising diffusion models \citep{SohlDickstein2015,Ho2020} provide another generative approach, with applications to super-resolving 2D \citep{Schanz2024} and 3D \citep{Rouhiainen2024} density fields.

The series of papers to which this work belongs has progressively extended a GAN-based SR framework that operates directly on $N$-body particles, providing outputs that can be used for halo identification, merger-tree construction, and statistical analysis. \citet{Li2021} introduced a Wasserstein-GAN model for super-resolving particle displacement fields at fixed cosmology (Paper~I). \citet{Ni2021} extended the framework to the full six-dimensional phase space, comprising displacement and velocity, and validated substructure, velocity, and higher-order clustering statistics (Paper~II). \citet{Zhang2023} used a StyleGAN2-adapted architecture \citep{Karras2019} in which the cosmic scale factor $a$ is injected as a style parameter, enabling SR across redshifts and tests of temporal consistency through halo merger trees (Paper~III). \citet{Zhang2025} added a complementary deterministic emulator conditioned on the HR initial condition (HRIC), using an SR realization and the HRIC to predict the corresponding HR outcome (Paper~IV); the same two-stage model was also applied to cosmological hydrodynamics \citep{hydroEmu}. The same SR framework has also been applied to simulations with initial power spectra modified to represent fuzzy dark matter \citep{Sipp2023}.

This paper extends the particle-based SR framework to cosmology-aware
generation. We train across the Quijote Latin-hypercube (LH) suite \citep{VillaescusaNavarro2020}, in which the five $\Lambda$CDM parameters $(\Omega_m, \Omega_b, h, n_s, \sigma_8)$ are varied simultaneously over a broad volume of parameter space. We employ these parameters as style inputs alongside the scale factor, extending the style vector from $\mathbf{s}_{\mathrm{style}} = a$ in Paper III to $\mathbf{s}_{\mathrm{style}} = (a, \Omega_m, \Omega_b, h, n_s, \sigma_8)$. The StyleGAN2 weight modulation layer in each convolutional block naturally absorbs additional style components, allowing a single model to learn the LR to HR mapping across the sampled cosmologies (Section~\ref{sec:method}). The training and evaluation presented here are at $z=0$, with $a=1$. The field-level emulators of \citet{Jamieson2023,Jamieson2024} learn evolution from linear displacement fields, with cosmological information entering through the initial fields and explicit conditioning parameters. Our generator learns a stochastic LR-to-HR map conditioned on evolved LR phase-space fields and the five cosmological parameters listed above. Related work has also conditioned particle SR on several cosmological parameters: \citet{Fremstad2026} train a GAN that super-resolves the displacement field as a function of $(\Omega_k,\Omega_m,\sigma_8,h)$ and redshift, with velocities estimated from the time dependence of the generated displacements. Our model jointly generates displacements and velocities, and we assess the resulting particle realizations through matter clustering and halo-population statistics.

We train the resulting model on 100 LR--HR simulation pairs sampled from the Quijote LH suite: $1\,h^{-1}\,\mathrm{Gpc}$ boxes pairing $512^3$ LR against $1024^3$ HR particles. The pairs are drawn from 100 distinct LH cosmologies, each with its own random seed at $z=0$. Because the LH set varies all five parameters simultaneously, the training data sample their joint effects on structure formation. For evaluation we hold out ten further LH cosmologies, disjoint from the SR training set, comprising a near-fiducial reference and nine additional cosmologies within the Quijote parameter ranges.

We validate the model using the matter power spectrum, halo mass function, subhalo mass function, and mean satellite occupation. The analysis measures the uncalibrated SR errors and tests the transfer of corrections from the reference cosmology to the other nine test cosmologies. We also quantify the parameter-level impact of the power-spectrum residuals by passing matched HR and SR spectra through the same neural posterior estimator trained on HR spectra. This comparison measures the parameter shifts associated with substituting SR predictions for HR inputs.

The paper is structured as follows. Section~\ref{sec:method} describes the dataset, model architecture, loss function, and training procedure. Section~\ref{sec:results} presents validation results across all ten test cosmologies, including the calibration and posterior-estimator tests. In Section~\ref{sec:discussion} we discuss our findings and summarize our results.
\section{Method}
\label{sec:method}

\begin{figure*}[t]
    \centering
    \includegraphics[width=\textwidth]{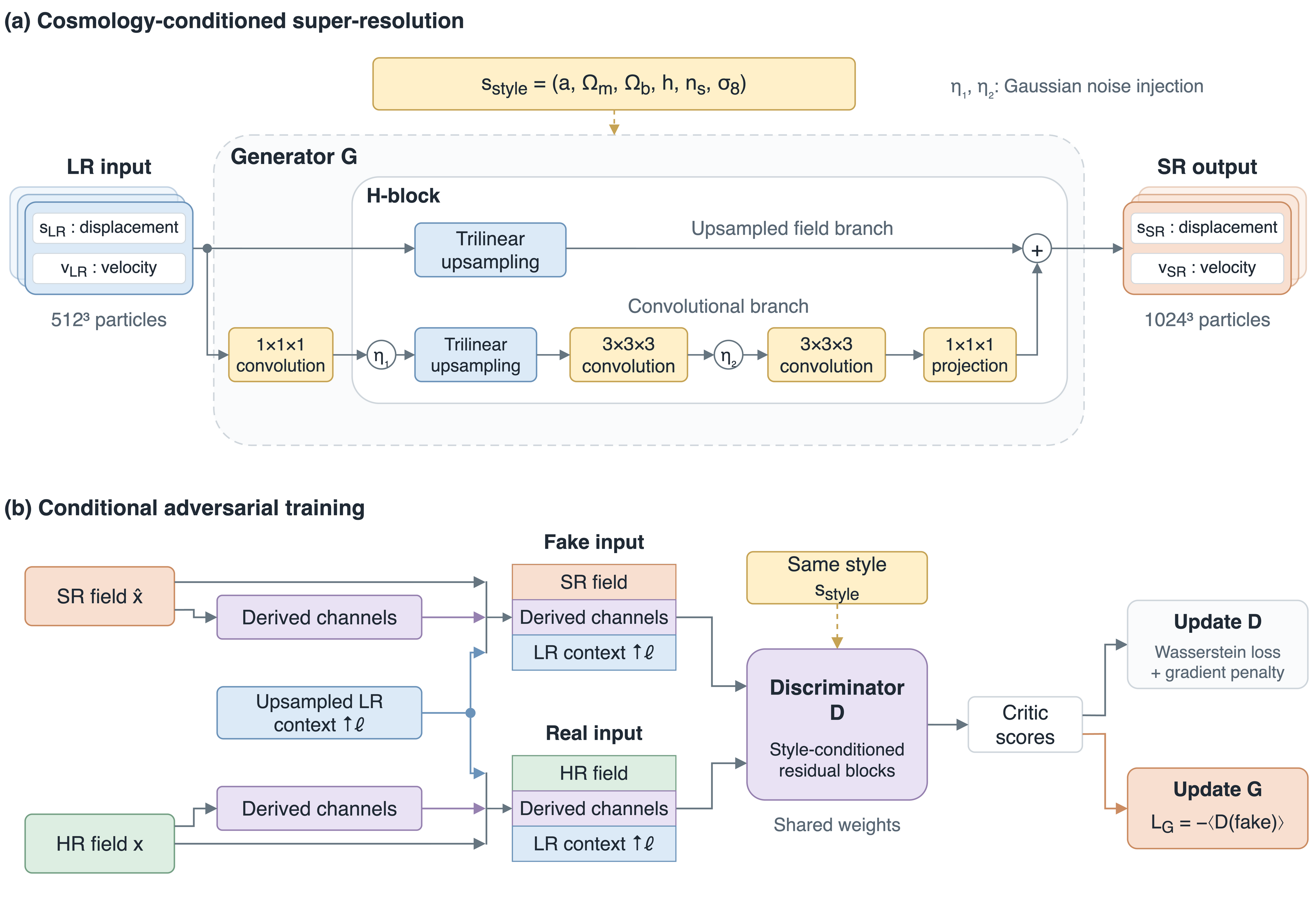}
    \caption{Cosmology-conditioned SR architecture and adversarial training.
    \textit{(a)} The generator architecture, conditioned on the style parameter; training operates on chunks.
    \textit{(b)} The candidate SR or HR field is concatenated with the common upsampled LR context and derived channels. The same style-conditioned discriminator evaluates both inputs and supplies the critic scores for the Wasserstein distance objectives.}
    \label{fig:architecture}
\end{figure*}

Figure~\ref{fig:architecture} summarizes our cosmology-conditioned
SR model and its adversarial training. The generator produces SR
particle displacements and velocities from LR fields, while the
discriminator evaluates the generated and HR fields using the same
LR context and cosmological conditioning.

\subsection{Dataset}
\label{sec:data}

The training and test data are drawn from the public Quijote simulation suite\footnote{\url{https://quijote-simulations.readthedocs.io}} \citep{VillaescusaNavarro2020},  a collection of cosmological $N$-body simulations designed for studies of cosmological information and machine-learning applications. The simulations were run with the \textsc{TreePM} code \textsc{Gadget-III}, an updated version of \textsc{Gadget-II} \citep{Springel2005}. Each box is $1\,h^{-1}\,\mathrm{Gpc}$ on a side. Initial conditions were generated at $z=127$ using linear matter power spectra and transfer functions from \textsc{CAMB} \citep{Lewis2000} and second-order Lagrangian perturbation theory \citep[2LPT;][]{Crocce2006}.

We draw both the training and the test data from the Quijote \emph{Latin-hypercube} (LH) set, in which the five $\Lambda$CDM parameters are varied \emph{simultaneously} over the ranges:
\begin{align}
\Omega_m  &\in [0.10,\ 0.50], \nonumber\\
\Omega_b  &\in [0.03,\ 0.07], \nonumber\\
h         &\in [0.50,\ 0.90], \nonumber\\
n_s       &\in [0.80,\ 1.20], \nonumber\\
\sigma_8  &\in [0.60,\ 1.00], \label{eq:lh}
\end{align}
with the neutrino mass and dark-energy equation of state held fixed at $M_\nu = 0$ and $w = -1$. We pair the standard LH simulations, containing $512^3$ particles, with the corresponding high-resolution LH simulations, containing $1024^3$ particles. We refer to these as the low-resolution (LR) and high-resolution (HR) samples, respectively. This gives a fixed SR factor of $2$ along each spatial axis, or $8$ in particle count.
The LR and HR simulations with the same LH identifier share the initial random seed and the phases of the modes resolved by both simulations; the HR initial conditions contain additional short-wavelength modes. Different LH cosmologies use independent initial random seeds.

The training set comprises 100 LR--HR pairs from 100 distinct LH cosmologies at $z=0$. These form a subset of the full LH design in Eq.~(\ref{eq:lh}). For evaluation, we use 10 further LH cosmologies that are excluded from SR training: a near-fiducial reference and nine additional points. We label the latter by their dominant parameter offset relative to the reference: $\Omega_m^{\pm}$, $\Omega_b^{\pm}$, $\sigma_8^{\pm}$, $n_s^{\pm}$, and $h^{-}$. In every test cosmology, all five parameters differ from the reference; the labels identify the individual test cases.

Each snapshot is represented by a six-channel field on the regular Lagrangian grid: three components of the displacement $\mathbf{s}$ and three components of the velocity $\mathbf{v}$. If $\mathbf{q}$ is a particle's Lagrangian grid coordinate, its Eulerian position is $\mathbf{x}=\mathbf{q}+\mathbf{s}(\mathbf{q})$, with positions wrapped periodically onto the simulation box. At training time, the fields are divided into cubical chunks with side lengths of $72$ LR grid cells and $144$ HR grid cells. LR chunks are padded by $3$ cells on each face, and HR targets by $4$ cells, using periodic boundaries. Data augmentation is applied at load time through axis-aligned reflections, permutations of the spatial axes, and periodic translations of the crop window by up to half the chunk side length.

\subsection{Model architecture}
\label{sec:architecture}

The model follows the StyleGAN2-adapted generator and discriminator framework of Paper~III \citep{Zhang2023,Karras2019}. We extend the style input to include the five cosmological parameters in Eq.~(\ref{eq:lh}). The generator maps the LR displacement and velocity fields to an SR field with the same six-channel representation as the HR target. The discriminator receives the LR context together with the candidate HR-resolution field and derived features, as described below.

\subsubsection{Style-conditioned generator}
\label{sec:generator}

The generator $G$ uses the H-block design of our earlier SR models \citep{Li2021,Ni2021,Zhang2023}, combining trilinear upsampling, style-modulated convolutions, stochastic noise injection, and skip connections. The total spatial upsampling factor in this work is $2$, achieved by a single H-block that doubles the grid resolution along each spatial axis, as shown in Figure~\ref{fig:architecture}(a).  Explicit cosmological parameter conditioning enters through the style vector
\begin{equation}
\mathbf{s}_{\mathrm{style}}
 = \bigl(a,\,\Omega_m,\,\Omega_b,\,h,\,n_s,\,\sigma_8\bigr),
\label{eq:style}
\end{equation}
which is passed to a learned linear mapping at each modulated convolution. All training and evaluation in this work use $z=0$, so $a=1$ throughout.

The convolutions use StyleGAN2 weight modulation and demodulation \citep{Karras2019}. For input channel $i$, output channel $j$, and kernel position $k$, the mapped style supplies a channel scale $m_i$:
\begin{align}
w'_{ijk} &= m_i(\mathbf{s}_{\mathrm{style}})\,w_{ijk},
\nonumber\\
w''_{ijk} &= \frac{w'_{ijk}}
 {\sqrt{\sum_{i,k}(w'_{ijk})^2+\epsilon}},
\label{eq:style_modulation}
\end{align}
where $\epsilon$ is a small positive constant. Demodulation normalizes the modulated weights separately for each output channel, allowing the convolutional filters to depend on the supplied cosmology while controlling their overall scale. Gaussian noise fields, multiplied by learned channel amplitudes, are added to the intermediate features to allow stochastic variation at fixed LR input and cosmology.

\subsubsection{Discriminator}
\label{sec:discriminator}
The discriminator $D$ uses style-conditioned residual convolutional blocks following Paper~III \citep{Zhang2023}. Each block combines a convolutional branch with a skip connection and downsamples the spatial features by a factor of $2$. The style vector in Eq.~(\ref{eq:style}) is supplied to the modulated layers, conditioning the critic on the input cosmology \citep{Mirza2014}.

The discriminator input combines the six upsampled LR displacement and velocity channels, the six channels of the SR output or HR target, and Eulerian-density features derived from the displacement field through Cloud-in-Cell (CIC) mass assignment (Figure~\ref{fig:architecture}(b)). The LR channels provide the spatial context against which the candidate HR-resolution field is evaluated.

\subsection{Loss function}
\label{sec:loss}
The LR simulations do not specify the additional short-wavelength modes present in the HR simulations. We therefore model SR as a conditional stochastic mapping, with noise inputs allowing different outputs for the same LR realization and cosmology. We train with an adversarial objective and use the paired HR field as a real sample for the discriminator. A per-voxel reconstruction loss is not included in the optimization objective. This differs from the deterministic emulation task of Paper~IV \citep{Zhang2025}, which also supplies the high-resolution initial conditions to identify the target realization.

Let $\ell$ and $x$ denote the paired LR and HR fields in the numerical representation used for training, and let $\boldsymbol{\eta}$ denote the collection of Gaussian noise inputs. The generated field is
\begin{equation}
\hat{x}=G(\ell,\mathbf{s}_{\mathrm{style}},\boldsymbol{\eta}).
\label{eq:conditional_generator}
\end{equation}
We write $\mathbf{c}=(\ell,\mathbf{s}_{\mathrm{style}})$ for the fixed conditioning information and $D(x;\mathbf{c})$ for the critic score after assembling the candidate field, LR context, and derived features described in Section~\ref{sec:discriminator}.

We minimize the Wasserstein-GAN losses \citep{Arjovsky2017} with a gradient penalty \citep{Gulrajani2017}:
\begin{align}
\mathcal{L}_D
 &= \mathbb{E}\!\left[D(\hat{x};\mathbf{c})-D(x;\mathbf{c})\right]
 \nonumber\\
 &\quad +\lambda_{\mathrm{GP}}\,
 \mathbb{E}\!\left[
  \left(\left\lVert\nabla_{\tilde{x}}
  D(\tilde{x};\mathbf{c})\right\rVert_2-1\right)^2
 \right],
\label{eq:LD}\\
\mathcal{L}_G
 &= -\mathbb{E}\!\left[D(\hat{x};\mathbf{c})\right].
\label{eq:LG}
\end{align}
The expectations are over paired training samples, the generator noise, and, for the penalty, an independent interpolation coefficient $t\sim\mathcal{U}[0,1]$. We construct $\tilde{x}=t x+(1-t)\hat{x}$ while holding $\mathbf{c}$ fixed. The gradient is taken with respect to this interpolated six-channel field; the derived discriminator features are recomputed from $\tilde{x}$ before evaluating the critic. We set $\lambda_{\mathrm{GP}}=10$.

We evaluate the gradient penalty once every 16 discriminator updates, multiplying its coefficient by 16 on those updates, using the lazy regularization strategy of \citet{Karras2019}. This compensates for the lower evaluation frequency in the regularizer's average coefficient. The paired mean-squared error is recorded as a diagnostic and does not contribute gradients to either network during adversarial training.

\subsection{Optimization and training procedure}
\label{sec:training}

We optimize both networks with Adam \citep{KingmaBa2015}, using $\beta_1=0$ and $\beta_2=0.99$. The learning rates are $\eta_G=10^{-5}$ for the generator and $\eta_D=2\times10^{-5}$ for the discriminator. Training uses one LR--HR chunk pair per GPU in each batch, which is the largest batch size we can fit on our GPU.

\section{Results}
\label{sec:results}
We evaluate the cosmology-aware super-resolution model described in Section~\ref{sec:method} on the 10 Quijote test cosmologies introduced in Section~\ref{sec:data}. We compare the projected particle distributions, matter power spectra, and halo and subhalo statistics, and examine the effect of SR--HR differences on cosmological parameter inference.

\subsection{Visual Comparison Across Cosmologies}
\label{subsec:visual}

\begin{figure*}
\centering
\includegraphics[width=\textwidth]{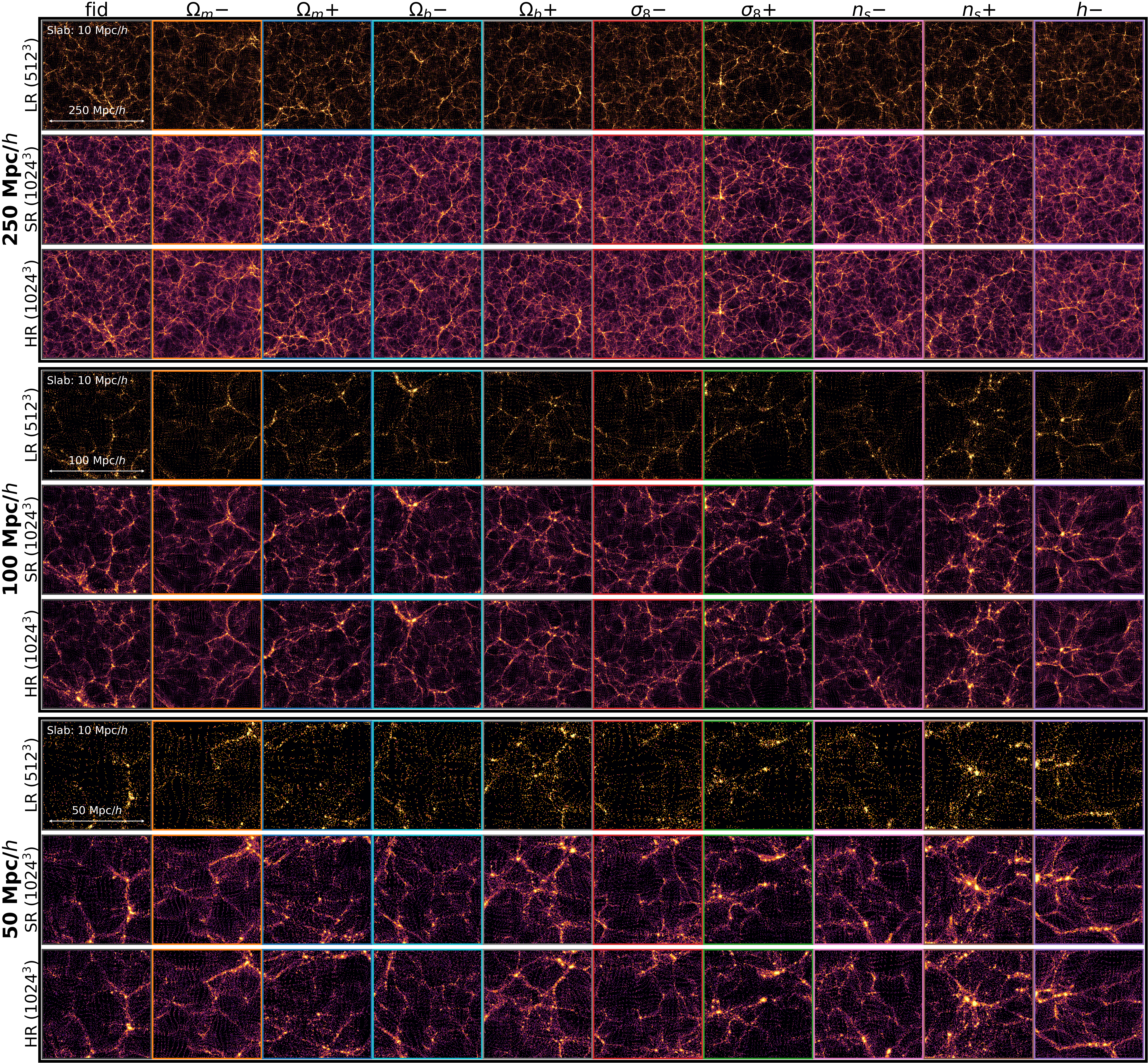}
\caption{Projected particle distributions at $z=0$ for the 10 Quijote LH test cosmologies. The three groups of rows show fields of view with side lengths of $250\,\mpch$, $100\,\mpch$, and $50\,\mpch$, each projected through a slab of thickness $10\,\mpch$. All three fields of view are centered on the geometric center of the simulation box, with the same projection center used for LR, SR, and HR. Within each group, the rows show the LR input (top), SR output (middle), and HR target (bottom). The $512^3$ and $1024^3$ particle counts refer to the full simulation boxes. LR particles are assigned eight times the rendering weight of HR particles to account for the difference in particle number. The columns show the near-fiducial reference (fid), followed by the nine test cases labeled $\Omega_m^{-}$, $\Omega_m^{+}$, $\Omega_b^{-}$, $\Omega_b^{+}$, $\sigma_8^{-}$, $\sigma_8^{+}$, $n_s^{-}$, $n_s^{+}$, and $h^{-}$. The labels identify the dominant parameter offset; all five parameters differ from the reference in each non-reference case. Different columns use independent initial random seeds. Colored borders identify the cosmologies and match the line colors in the statistical comparisons.}
\label{fig:visual}
\end{figure*}

Figure~\ref{fig:visual} compares the LR, SR, and HR particle distributions rendered with \texttt{gaepsi2}\footnote{\url{https://github.com/rainwoodman/gaepsi2}} across the test cosmologies. Within each column, the LR and HR simulations share the initial-condition phases of their common modes, and the SR field is generated from the corresponding LR input. The aligned projections therefore allow the SR and HR morphology to be compared within the same large-scale realization.

At the $250\,\mpch$ level, the SR and HR maps are visually indistinguishable in every column. The SR maps reproduce the prominent nodes, the connected filaments, and the void regions seen in the HR projections. The LR inputs trace the main structures with a coarser particle sampling: their nodes are diffuse, the thinnest filaments are washed out, and the small-scale texture inside collapsed regions is absent. The SR model restores all of these features starting from the coarse LR field.

The $100\,\mpch$ and $50\,\mpch$ panels show the comparison in more detail. Particle discreteness becomes more apparent in the LR maps, particularly along filaments and in underdense regions: the LR input reduces to isolated bright particles, filaments are broken into dotted chains, and the Lagrangian particle lattice is visible. The SR output reveals finer structure along the filaments and around the dense regions, with a morphology similar to the HR maps. 

Across the columns, the SR--HR correspondence is visible in regions with different filament configurations and concentrations of dense structures. Differences between columns reflect both the jointly varying cosmological parameters and the independent initial realizations. The quantitative comparisons of the matter power spectrum and halo statistics in Sections~\ref{subsec:pk} and~\ref{subsec:halos} assess the statistical agreement across these test cosmologies.
\subsection{Matter Power Spectrum}
\label{subsec:pk}

\begin{figure*}
\centering
\includegraphics[width=0.93\textwidth]{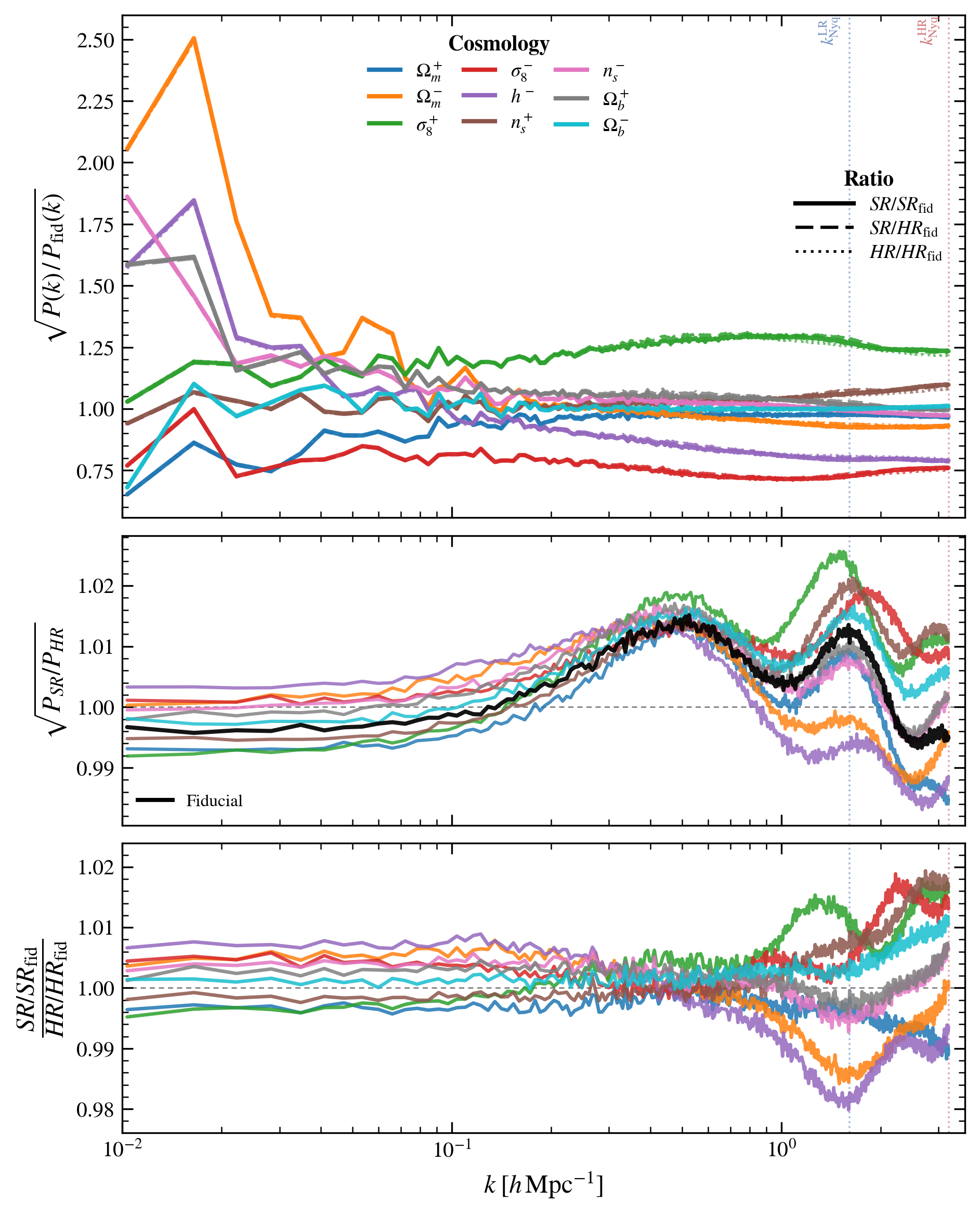}
\caption{Matter power-spectrum comparison across the nine non-fiducial Quijote test cosmologies, with each color denoting one cosmology.
\emph{Top:} square roots of the power-spectrum ratios relative to the fiducial reference. Solid curves show
$\sqrt{P_{\mathrm{SR}}(k;\boldsymbol{\theta})/
P_{\mathrm{SR}}(k;\boldsymbol{\theta}_{\mathrm{fid}})}$,
dotted curves show
$\sqrt{P_{\mathrm{HR}}(k;\boldsymbol{\theta})/
P_{\mathrm{HR}}(k;\boldsymbol{\theta}_{\mathrm{fid}})}$,
and dashed curves show
$\sqrt{P_{\mathrm{SR}}(k;\boldsymbol{\theta})/
P_{\mathrm{HR}}(k;\boldsymbol{\theta}_{\mathrm{fid}})}$.
\emph{Middle:} direct SR-to-HR accuracy, shown as the amplitude ratio
$A=\sqrt{P_{\mathrm{SR}}/P_{\mathrm{HR}}}$ at each cosmology;
the black curve denotes the fiducial reference.
\emph{Bottom:} the SR-to-HR power-spectrum ratio after removing the
scale-dependent model bias measured at the fiducial cosmology,
$Q=(P_{\mathrm{SR}}/P_{\mathrm{HR}})/
(P_{\mathrm{SR,fid}}/P_{\mathrm{HR,fid}})$
(Equation~\ref{eq:pk_Q}).
Departures from unity measure the residual error after this correction.
Horizontal gray dashed lines mark unity. The blue and red
vertical dotted lines mark the LR and HR particle-grid Nyquist
wavenumbers, respectively.}
\label{fig:pk_all}
\end{figure*}

We compute the matter power spectrum $P(k)$ with \texttt{nbodykit} \citep{Hand2018}. For each realization, we assign the particles in the $1\,h^{-1}\mathrm{Gpc}$ box to a mesh with $N_{\rm mesh}=1024$ cells per side using Cloud-in-Cell (CIC) mass assignment. We Fourier-transform the density contrast and average the power in spherical shells in $|\boldsymbol{k}|$. CIC window compensation and interlacing are disabled. The mesh spans wavenumbers from the fundamental mode, $k_{\rm f}=2\pi/L\simeq6.3\times10^{-3}\,h\,\mathrm{Mpc}^{-1}$, to its Nyquist wavenumber, $k_{\rm Nyq}^{\rm mesh}=\pi N_{\rm mesh}/L \simeq3.2\,h\,\mathrm{Mpc}^{-1}$. All HR targets and SR outputs are evaluated using the same mesh, mass-assignment scheme, and $k$-binning.

Figure~\ref{fig:pk_all} compares the cosmological response, the direct SR-to-HR accuracy at each cosmology, and the residual error after removing the model bias measured at the fiducial cosmology. For a test cosmology
$\boldsymbol{\theta}$, we define
\begin{equation}
\mathcal{R}_{X|Y}(k;\boldsymbol{\theta})
\equiv
\sqrt{
\frac{P_X(k;\boldsymbol{\theta})}
{P_Y(k;\boldsymbol{\theta}_{\rm fid})}
},
\label{eq:pk_response}
\end{equation}
where $X,Y\in\{\mathrm{SR},\mathrm{HR}\}$ specify the test and reference spectra, respectively.

The top panel shows $\mathcal{R}_{\rm SR|SR}$ as solid curves, $\mathcal{R}_{\rm SR|HR}$ as dashed curves, and $\mathcal{R}_{\rm HR|HR}$ as dotted curves. The solid and dotted curves are the internally normalized SR and HR responses. A multiplicative SR-to-HR bias $b(k)$ common to the reference and test cosmologies cancels from $\mathcal{R}_{\rm SR|SR}$. The agreement between the solid and dotted curves shows that the SR spectra follow the changes in the corresponding HR spectra across the test set, including their scale dependence. The dashed curves also retain the fiducial SR-to-HR offset: at each $k$, their ratio to the solid curves equals $\sqrt{P_{\rm SR}(k;\boldsymbol{\theta}_{\rm fid})/ P_{\rm HR}(k;\boldsymbol{\theta}_{\rm fid})}$. Because different test cosmologies use independent initial seeds, the displayed changes include both cosmological and realization differences.

The middle panel measures the direct SR-to-HR accuracy at each cosmology, before any bias correction, through the amplitude  ratio
\begin{equation}
A(k;\boldsymbol{\theta})
\equiv
\sqrt{
\frac{P_{\rm SR}(k;\boldsymbol{\theta})}
{P_{\rm HR}(k;\boldsymbol{\theta})}
}.
\label{eq:pk_A}
\end{equation}

The black curve denotes the fiducial reference, and the colored curves denote the nine non-fiducial cosmologies. The deviations from unity are small but scale dependent. The spectra agree closely on large scales and show a common excess of SR power at intermediate wavenumbers, followed by a larger spread among the test cases at high $k$. We express the fractional error in power as

\begin{equation}
\epsilon_{\rm abs}(k;\boldsymbol{\theta})
\equiv
A^2(k;\boldsymbol{\theta})-1,
\label{eq:pk_eps_abs}
\end{equation}

which gives $\epsilon_{\rm abs}\simeq2(A-1)$ for small amplitude differences. At the fiducial cosmology, the maximum of $|\epsilon_{\rm abs}|$ is 3.1\% over $0.1<k<3\,h\,\mathrm{Mpc}^{-1}$, with a mean signed error of +0.9\%. Across all ten cosmologies, including the fiducial case, the mean of the per-cosmology maximum absolute power errors over this interval is 3.6\%.

To separate the model bias measured at the fiducial cosmology from the remaining differences across the test cosmologies, we define the scale-dependent fiducial bias as
\begin{equation}
b_{\mathrm{fid}}(k)
\equiv
\frac{
P_{\mathrm{SR}}(k;\boldsymbol{\theta}_{\mathrm{fid}})
}{
P_{\mathrm{HR}}(k;\boldsymbol{\theta}_{\mathrm{fid}})
}.
\label{eq:pk_fid_bias}
\end{equation}
We apply the same correction factor $1/b_{\rm fid}(k)$ to each SR spectrum. The bottom panel compares the corrected spectrum with its corresponding HR target through the calibrated power-spectrum ratio
\begin{equation}
\begin{aligned}
Q(k;\boldsymbol{\theta})
&\equiv
\frac{
P_{\mathrm{SR}}(k;\boldsymbol{\theta})/
P_{\mathrm{HR}}(k;\boldsymbol{\theta})
}{
P_{\mathrm{SR}}(k;\boldsymbol{\theta}_{\mathrm{fid}})/
P_{\mathrm{HR}}(k;\boldsymbol{\theta}_{\mathrm{fid}})
}
\\
&=
\frac{1}{b_{\mathrm{fid}}(k)}
\frac{P_{\mathrm{SR}}(k;\boldsymbol{\theta})}
{P_{\mathrm{HR}}(k;\boldsymbol{\theta})}.
\end{aligned}
\label{eq:pk_Q}
\end{equation}
This removes the fiducial model bias from the direct SR-to-HR comparison. If that bias is shared by the reference and test cosmologies, it cancels exactly and $Q=1$. Departures from unity measure the residual error after the fiducial correction and thus test how well a calibration determined at one cosmology transfers to the other test cases.

The fractional residual in power is
\begin{equation}
\epsilon_{\rm corr}(k;\boldsymbol{\theta})
\equiv
Q(k;\boldsymbol{\theta})-1.
\label{eq:pk_eps_corr}
\end{equation}
The bottom-panel curves stay within about $2\%$ of unity across the nine non-fiducial cosmologies. The fiducial correction accounts for the common intermediate-scale offset, while the remaining differences vary across the test cases.

The agreement persists through the LR particle Nyquist frequency, $k_{\rm Nyq}^{\rm LR}\simeq1.6\,h\,\mathrm{Mpc}^{-1}$, and into the interval between the LR and HR particle Nyquist frequencies. In this regime, the LR field no longer contains independently resolved Fourier modes, and the SR model generates the unresolved small-scale structure from the resolved LR field, the stochastic input, and the cosmological conditioning. The agreement with HR in this regime shows that the generated particle distribution reproduces the small-scale power and carries the correct cosmological dependence.

Together, these results show that the cosmology-conditioned SR model reproduces the measured matter power spectra to within a few per cent at the tested cosmologies. The internally normalized SR spectra also follow the corresponding HR variations. The fiducial-calibrated ratios quantify the remaining SR-to-HR error across the test set. This agreement is obtained without an explicit power-spectrum term in the training loss (Section~\ref{sec:loss}).
\subsection{Halo and Subhalo Statistics}
\label{subsec:halos}

The matter power spectrum measures the clustering of the density field, while halo and subhalo statistics test the abundance of collapsed structures and their substructure formed by nonlinear gravitational evolution. Because the SR model outputs the full particle phase space, its predictions can be processed with the same halo-finding pipeline as the HR simulations \citep{Ni2021}.

Particles separated by less than $b=0.2$ times the mean inter-particle separation are connected into friends-of-friends (FoF) groups \citep{Davis85}. Within each FoF group, \textsc{SubFind} identifies locally overdense structures and iteratively removes gravitationally unbound particles \citep{Springel2001subfind}. We use the total FoF group mass for the halo mass function and the bound \textsc{SubFind} mass for the subhalo statistics below. Satellite subhalos are defined as bound substructures distinct from the central subhalo of their FoF host. Centrals are excluded from both the subhalo mass function and satellite occupation. We retain only objects containing at least 100 particles. Within each cosmology, the matched SR and HR fields have the same particle mass and are analyzed with the same FoF/\textsc{SubFind} pipeline and parameters. At a fixed particle number threshold, the corresponding mass threshold varies across cosmologies through the particle mass.

Each cosmology is represented by a single matched SR--HR pair. Their shared initial-condition phases cancel sample variance in the paired comparisons. The shaded regions in the figures show Poisson counting uncertainties, $\sigma_N=\sqrt{N}$, propagated to the ratios in the Gaussian limit.

As in the power spectrum comparison, the three panels separate the cosmological response, direct SR-to-HR accuracy, and residual error after removing the model bias measured at the fiducial cosmology. For a statistic $S$ in a given mass bin, the top panels show $S_X(\boldsymbol{\theta})/S_Y(\boldsymbol{\theta}_{\rm fid})$: solid curves have $X=Y=\mathrm{SR}$, dotted curves have $X=Y=\mathrm{HR}$, and dashed curves have $X=\mathrm{SR}$ and $Y=\mathrm{HR}$. All ratios in this section are formed directly from the measured statistic, without square roots. The middle panels show the direct SR-to-HR ratio
\begin{equation}
B_S(\boldsymbol{\theta})
\equiv
\frac{S_{\rm SR}(\boldsymbol{\theta})}
     {S_{\rm HR}(\boldsymbol{\theta})},
\label{eq:halo_B}
\end{equation}
and the bottom panels show the ratio after the fiducial bias correction,
\begin{equation}
\begin{aligned}
Q_S(\boldsymbol{\theta})
&\equiv
\frac{B_S(\boldsymbol{\theta})}
     {B_S(\boldsymbol{\theta}_{\rm fid})}
\\
&=
\frac{S_{\rm SR}(\boldsymbol{\theta})/
      S_{\rm SR}(\boldsymbol{\theta}_{\rm fid})}
     {S_{\rm HR}(\boldsymbol{\theta})/
      S_{\rm HR}(\boldsymbol{\theta}_{\rm fid})}.
\end{aligned}
\label{eq:halo_Q}
\end{equation}
The calibration is determined separately for each statistic as a function of its mass variable and applied to all nine non-fiducial cosmologies. A bias shared by the reference and test cosmologies cancels, giving $Q_S=1$. Departures from unity measure the residual error after this correction. In the middle panels, the black curve denotes the near-fiducial reference and the colored curves denote the nine non-fiducial cosmologies.

\subsubsection{Halo mass function}
\label{subsubsec:hmf}

\begin{figure*}[p]
\centering
\includegraphics[width=0.94\textwidth]{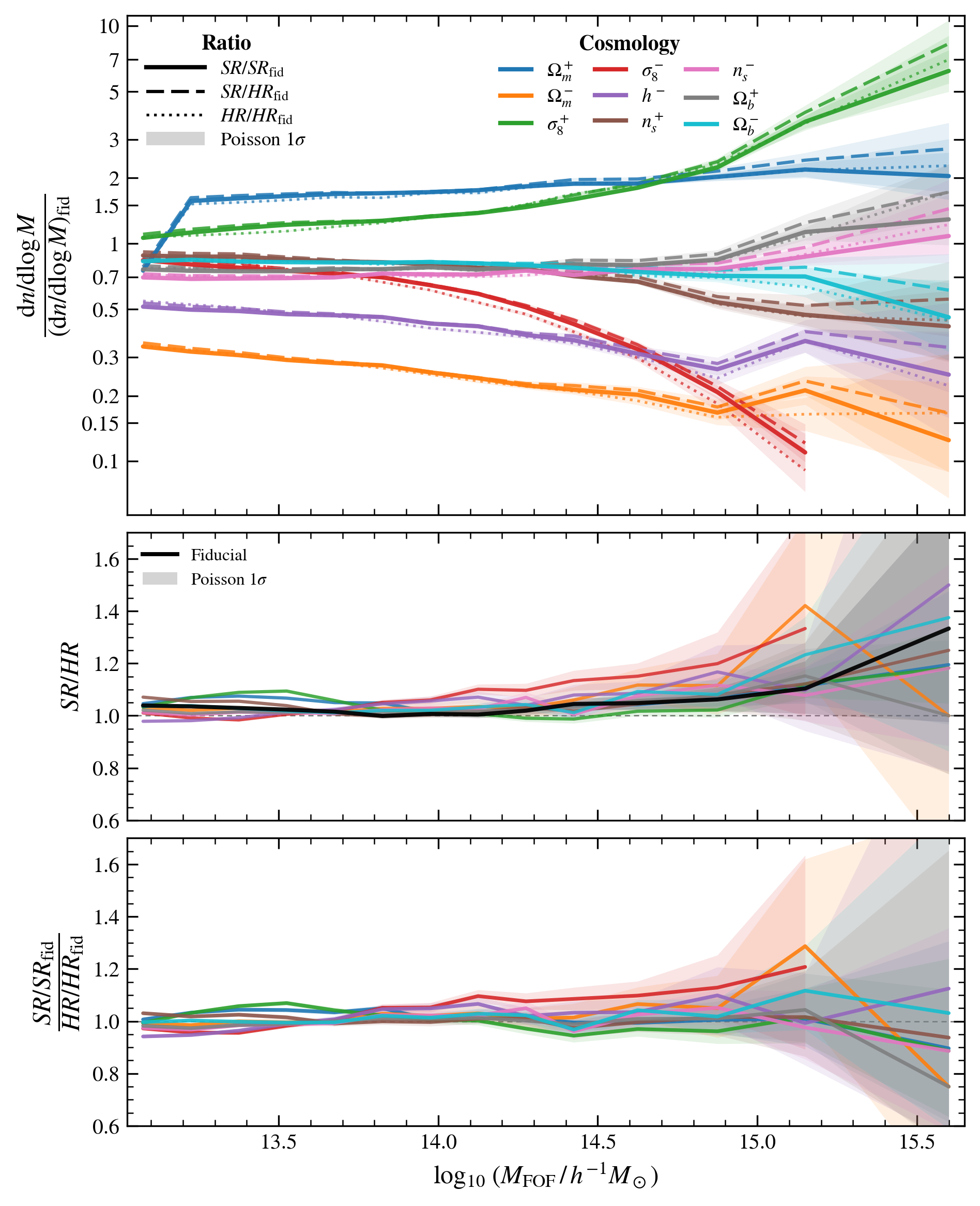}
\caption{FoF halo mass-function ratios at $z=0$ across the Quijote test cosmologies.
\textit{Top:} the abundance at each non-fiducial cosmology relative to the fiducial result. Solid curves show
$\Phi_{\rm SR}(\boldsymbol{\theta})/
 \Phi_{\rm SR}(\boldsymbol{\theta}_{\rm fid})$,
dashed curves show
$\Phi_{\rm SR}(\boldsymbol{\theta})/
 \Phi_{\rm HR}(\boldsymbol{\theta}_{\rm fid})$,
and dotted curves show
$\Phi_{\rm HR}(\boldsymbol{\theta})/
 \Phi_{\rm HR}(\boldsymbol{\theta}_{\rm fid})$,
where $\Phi=\mathrm{d}n/\mathrm{d}\log_{10}M_{\rm FoF}$.
\textit{Middle:} direct SR-to-HR accuracy for all ten cosmologies, shown as $B_{\rm HMF}$; the black curve denotes the fiducial result.
\textit{Bottom:} $Q_{\rm HMF}$, the SR-to-HR ratio after removing the mass-dependent model bias measured at the fiducial cosmology. Shaded regions show the propagated 1$\sigma$ Poisson uncertainties.}
\label{fig:hmf}
\end{figure*}

We define the FoF halo mass function as
\begin{equation}
\Phi_X(M;\boldsymbol{\theta})
\equiv
\frac{\mathrm{d}n_X(M;\boldsymbol{\theta})}
     {\mathrm{d}\log_{10}M_{\rm FoF}},
\label{eq:hmf_definition}
\end{equation}
where $n_X$ is the comoving number density and $X\in\{\mathrm{SR},\mathrm{HR}\}$. The mass bins widen toward the high-mass end, where halos are less abundant.

The solid and dotted curves in the top panel of Figure~\ref{fig:hmf} agree closely over much of the plotted mass range. The SR model reproduces the changes in halo abundance relative to the fiducial reference, including the enhanced high mass abundances in the cases labeled $\sigma_8^+$ and $\Omega_m^+$ and the suppressed abundances in the cases labeled $\sigma_8^-$ and $\Omega_m^-$. The largest contrasts occur in the high-mass tail, where the SR curves reproduce the broad ordering and mass dependence of the HR ratios.

The middle panel shows $B_{\rm HMF}$. At low and intermediate halo masses the SR and HR mass functions agree to within $15\%$ for $M_{\rm FoF} < 10^{14.8}\,h^{-1}M_\odot$ in all ten cosmologies, and to within $7\%$ at the fiducial cosmology. A mild excess of SR halos develops toward the cluster-mass regime, with the ratio generally increasing above unity as the halo mass increases, consistent in direction with the small-scale power excess of Section~\ref{subsec:pk} and possibly assisted by FoF bridging at $b=0.2$. In the highest mass bins the number of objects becomes small and the Poisson uncertainties broaden rapidly, producing the larger scatter among cosmologies seen in the panel.

The bottom panel shows $Q_{\rm HMF}$. Removing the fiducial bias accounts for the common upward trend with halo mass, and the corrected ratios remain close to unity over most of the plotted range. The larger departures occur in the highest-mass bins, where the uncertainty bands also broaden.

\subsubsection{Subhalo mass function}
\label{subsubsec:shmf}

\begin{figure*}[p]
\centering
\includegraphics[width=0.94\textwidth]{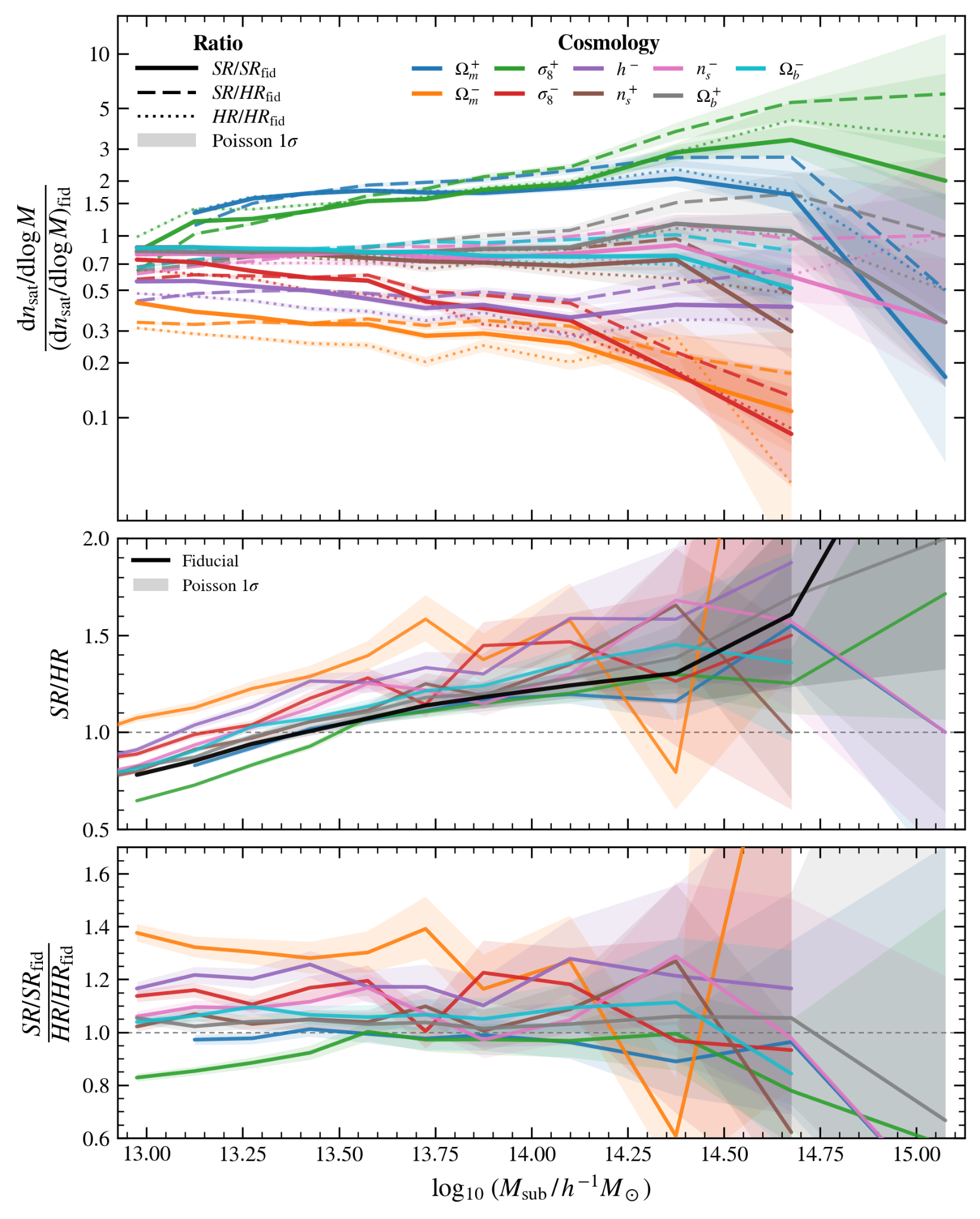}
\caption{Satellite subhalo mass-function ratios at $z=0$ across the Quijote test cosmologies; centrals are excluded.
\textit{Top:} solid curves show
$\Phi_{\rm sub,SR}(\boldsymbol{\theta})/
 \Phi_{\rm sub,SR}(\boldsymbol{\theta}_{\rm fid})$,
dashed curves show
$\Phi_{\rm sub,SR}(\boldsymbol{\theta})/
 \Phi_{\rm sub,HR}(\boldsymbol{\theta}_{\rm fid})$,
and dotted curves show
$\Phi_{\rm sub,HR}(\boldsymbol{\theta})/
 \Phi_{\rm sub,HR}(\boldsymbol{\theta}_{\rm fid})$,
where
$\Phi_{\rm sub}=\mathrm{d}n_{\rm sat}/\mathrm{d}\log_{10}M_{\rm sub}$.
\textit{Middle:} direct SR-to-HR accuracy for all ten cosmologies, shown as $B_{\rm sub}$; the black curve denotes the fiducial result.
\textit{Bottom:} $Q_{\rm sub}$, the ratio after removing the subhalo-mass-dependent model bias measured at the fiducial cosmology. Shaded regions show the propagated 1$\sigma$ Poisson uncertainties.}
\label{fig:shmf}
\end{figure*}

We define the subhalo mass function as
\begin{equation}
\Phi_{{\rm sub},X}(M_{\rm sub};\boldsymbol{\theta})
\equiv
\frac{
\mathrm{d}n_{{\rm sat},X}(M_{\rm sub};\boldsymbol{\theta})
}{
\mathrm{d}\log_{10}M_{\rm sub}
},
\quad
X\in\{\mathrm{SR},\mathrm{HR}\},
\label{eq:shmf_definition}
\end{equation}
where $n_{{\rm sat},X}$ is the comoving number density of satellite subhalos.

In the top panel of Figure~\ref{fig:shmf}, the SR model reproduces the ordering and overall mass dependence of the HR ratios across the test cosmologies. Differences become more pronounced toward the high mass end of the satellite population.
 
The middle panel shows $B_{\rm sub}$, with a systematic dependence on subhalo mass. At the low-mass end, the fiducial SR abundance is approximately $20\%$ below HR at $M_{\rm sub}\sim10^{13}\,h^{-1}M_\odot$. The ratio crosses unity near $10^{13.4}\,h^{-1}M_\odot$ and generally exceeds unity at higher mass. At $10^{14}$--$10^{14.5}\,h^{-1}M_\odot$, many of the SR to HR ratios lie around $1.2$--$1.5$, and the differences among cosmologies and the uncertainty bands also grow. The low-mass deficit sits close to the resolution threshold, where the bound \textsc{SubFind} mass is most sensitive to the unbinding step.

The bottom panel shows $Q_{\rm sub}$. The fiducial calibration removes much of the common mass-dependent tilt, bringing the responses closer to unity and revealing the remaining differences across cosmologies. These residuals at the low-mass end and above $\sim10^{14.5}\,h^{-1}M_\odot$ become more dispersed. The subhalo mass function is therefore less accurately reproduced than the FoF halo mass function: the broad cosmological response is captured, but the absolute abundance retains a systematic dependence on subhalo mass and a smaller cosmology-dependent residual after calibration.

\subsubsection{Mean occupation number}
\label{subsubsec:occupation}

\begin{figure*}[p]
\centering
\includegraphics[width=0.94\textwidth]{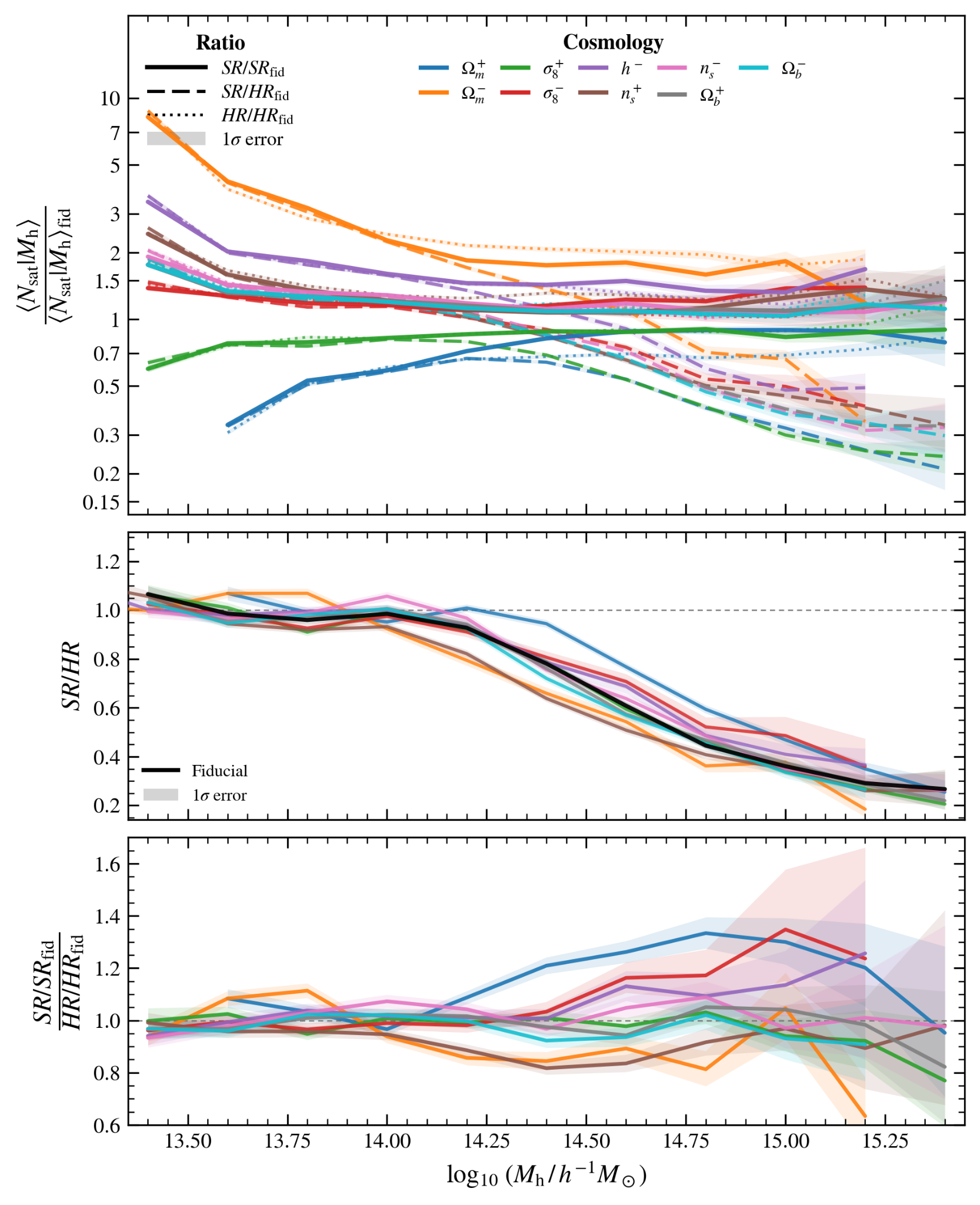}
\caption{Mean occupation ratios at $z=0$ across the Quijote test cosmologies.
\textit{Top:} solid curves show
$\mathcal{N}_{\rm SR}(\boldsymbol{\theta})/
 \mathcal{N}_{\rm SR}(\boldsymbol{\theta}_{\rm fid})$,
dashed curves show
$\mathcal{N}_{\rm SR}(\boldsymbol{\theta})/
 \mathcal{N}_{\rm HR}(\boldsymbol{\theta}_{\rm fid})$,
and dotted curves show
$\mathcal{N}_{\rm HR}(\boldsymbol{\theta})/
 \mathcal{N}_{\rm HR}(\boldsymbol{\theta}_{\rm fid})$,
where $\mathcal{N}(M_h)=\langle N_{\rm sat}\mid M_h\rangle$ and
$M_h=M_{\rm FoF}$.
\textit{Middle:} direct SR-to-HR accuracy for all ten cosmologies, shown as $B_{\rm occ}$; the black curve denotes the fiducial result.
\textit{Bottom:} $Q_{\rm occ}$, the ratio after removing the host mass dependent model bias measured at the fiducial cosmology. Shaded regions show the propagated 1$\sigma$ Poisson uncertainties.}
\label{fig:occupation}
\end{figure*}

The mean occupation number is the average number of resolved satellite subhalos in an FoF host of mass $M_h$:
\begin{equation}
\begin{aligned}
\mathcal{N}_X(M_h;\boldsymbol{\theta})
&\equiv
\left\langle N_{\rm sat}\mid M_h\right\rangle_X
\\
&=
\frac{1}{N_{{\rm host},X}(M_h)}
\sum_{i\,\in\,{\rm bin}(M_h)}N_{{\rm sat},i},
\quad
X\in\{\mathrm{SR},\mathrm{HR}\}
\end{aligned}
\label{eq:occupation_definition}
\end{equation}
where the sum runs over the $N_{{\rm host},X}(M_h)$ hosts in the mass bin centered on $M_h$. Unlike the global subhalo mass function, this statistic conditions the satellite abundance on host mass, and therefore tests the internal substructure of group and cluster mass halos directly.

The top panel of Figure~\ref{fig:occupation} shows substantial variation in mean occupation across the test cosmologies, particularly in the lower host-mass bins. The internally normalized SR ratios follow the broad HR trends, with more visible differences toward the high-mass end.
 
The middle panel shows $B_{\rm occ}$. In lower mass bins, the ratios lie within approximately 10\% of unity up to $M_h\sim10^{14.2}\,h^{-1}M_\odot$. The ratios then decline with increasing host mass. At $M_h\sim10^{15}\,h^{-1}M_\odot$ the SR mean occupation number is only about $30$--$50\%$ of the HR value, and the fraction decreases further in the most massive bins. The near fiducial and other nine test cases show the same trend,  so the deficit is a property of the architecture, not an artifact introduced by the cosmology conditioning. Similar satellite deficits were reported for the SR baseline in Paper~IV \citep{Zhang2025}, which discussed the finite spatial resolution of the discriminator's Eulerian density input as a possible contributor.

The bottom panel shows $Q_{\rm occ}$. Below $M_h\sim10^{14.2}\,h^{-1}M_\odot$ the corrected ratios stay within $11\%$ of unity. Above that the calibration removes the dominant suppression, moving the ratios from $0.2$--$0.5$ to $0.63$--$1.35$ centered near unity, with residuals at the tens-of-percent level for several cosmologies. Thus the calibration substantially improves the occupation comparison while retaining measurable differences among the test cases.
 
Together, these results show good agreement in FoF halo abundances, and for satellite level statistics, they carry a known, architecture-level deficit that is largely cosmology-independent, and the single-simulation calibration restores their cosmological response throughout the hypercube. The calibration demonstrates that a substantial part of the occupation deficit is shared across the test cosmologies. Accurately reproducing the uncalibrated satellite abundances and occupations in massive hosts remains a priority for further model development.
\subsection{Cosmological parameter inference}
\label{subsec:inference}

The comparisons above ask whether individual SR statistics match HR. We now assess the effect of the remaining power spectrum differences on cosmological parameter inference. If an SR spectrum replaces its matched HR spectrum as the input to the same posterior estimator, how much do the inferred parameters shift?

We construct a neural posterior estimator \citep[NPE;][]{Papamakarios2016} using the publicly released real-space matter power spectra from the Quijote HR Latin-hypercube suite \citep{VillaescusaNavarro2020}. From the full suite of 2,000 cosmologies, we exclude the 100 cosmologies used to train the SR model and the 10 used to test it, leaving 1,890 cosmologies for NPE fitting. The estimator maps a power spectrum data vector over $k_{\min}\leq k\leq2\,h\,\mathrm{Mpc}^{-1}$ at $z=0$ to an approximate joint posterior for $\boldsymbol{\theta}=(\Omega_m,\Omega_b,h,n_s,\sigma_8)$, with a uniform prior over the ranges in Eq.~(\ref{eq:lh}). We adopt $k_{\max} = 2\,h\,\mathrm{Mpc}^{-1}$, which lies below the Nyquist wavenumber of the $1024^3$ measurement mesh ($k^{\rm mesh}_{\rm Nyq} \simeq 3.2\,h\,\mathrm{Mpc}^{-1}$), while extending beyond the LR particle-grid Nyquist wavenumber ($k^{\rm LR}_{\rm Nyq} \simeq 1.6\,h\,\mathrm{Mpc}^{-1}$) so that the test includes scales generated by the SR model. SR spectra are not used to fit the posterior estimator.

The SR generator is stochastic by design (Section~\ref{sec:loss}), and the preceding comparisons assess its particle realizations through summary statistics. Here, $P(k)$ provides a common benchmark that relates the measured SR--HR differences to cosmological parameter sensitivity. The generated particle realizations also support other statistics, whose inference performance can be assessed separately.

For each matched pair $i$ and parameter $j$, we define the normalized posterior-mean shift as
\begin{equation}
\Delta_{ij}^{X}
\equiv
\frac{\hat{\theta}_{ij}^{X}-\hat{\theta}_{ij}^{\rm HR}}
     {\sigma_{ij}^{\rm HR}},
\qquad X\in\{\mathrm{SR},\mathrm{cal}\},
\label{eq:npe_shift}
\end{equation}
where $\hat{\theta}_{ij}^{X}$ is the marginal posterior mean obtained from input $X$, and $\sigma_{ij}^{\rm HR}$ is the corresponding HR posterior standard deviation. The label $\mathrm{cal}$ denotes the fiducial-calibrated SR spectrum defined below. The estimator is kept fixed for all inputs, so $\Delta_{ij}^{X}$ measures its response to replacing the HR spectrum. The shared initial phases of each matched pair reduce the realization variance in this comparison.

\begin{figure*}[p]
\centering
\includegraphics[width=0.94\textwidth]{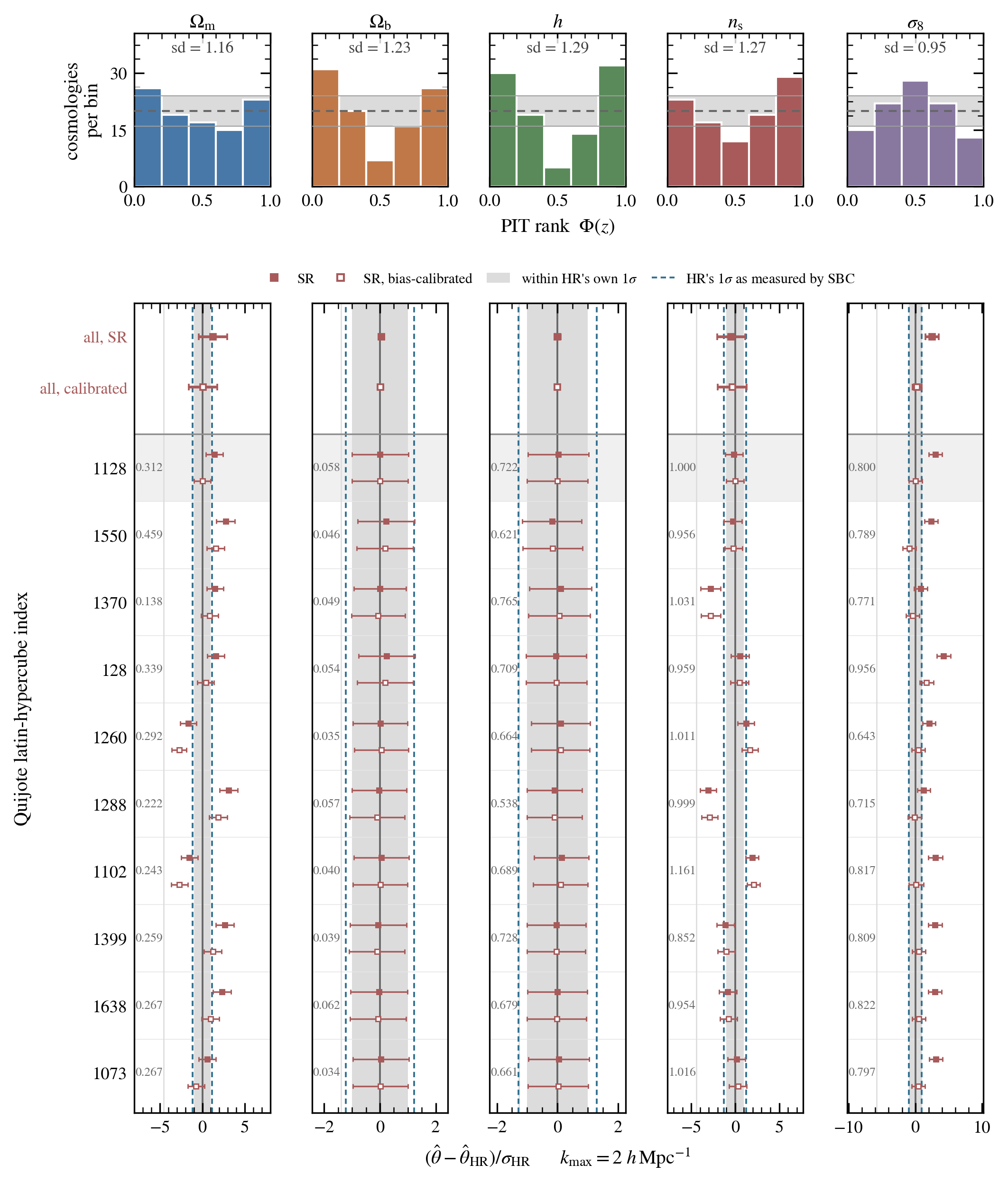}
\caption{Response of an HR-trained posterior estimator to SR modes up to $k_{\max}=2\,h\,\mathrm{Mpc}^{-1}$.
\textit{Upper panels:} marginal simulation-based calibration using 100 HR test cosmologies. Histograms show normalized posterior ranks of the true parameters, with the uniform expectation (dashed) and an approximate point-wise binomial $\pm1\sigma$ range (gray). Each panel reports the standard deviation of the normalized residual $z$ defined in Eq.~(\ref{eq:npe_sbc_z}).
\textit{Lower panels:} signed posterior-mean shifts $(\hat{\theta}^{X}-\hat{\theta}^{\rm HR})/\sigma^{\rm HR}$ for the ten matched pairs, using raw SR spectra (filled symbols) and fiducial-calibrated SR spectra (open symbols). Gray numbers give the true parameter values. The gray bands mark $\Delta=\pm1$, and the blue dashed lines mark $\Delta=\pm\mathrm{sd}(z)$ from the HR test sample. The two rows labeled ``all, SR'' and ``all, calibrated'' show the mean signed shift, $\overline{\Delta_j^{X}}=N_{\rm pair}^{-1}\sum_i\Delta_{ij}^{X}$ (symbols), and the standard deviation of $\Delta_{ij}^{X}$ (error bars) over the $N_{\rm pair}=9$ non-fiducial cosmologies. LH~1128 is the fiducial calibration case: its calibrated shift is zero by construction.}
\label{fig:inference}
\end{figure*}

We assess marginal posterior calibration using simulation-based calibration \citep[SBC;][]{Talts2018}. The upper panels of Figure~\ref{fig:inference} show normalized posterior ranks of the true parameters for 100 HR test cosmologies held out from NPE fitting. Uniform ranks are expected for calibrated posteriors under the assumed prior predictive distribution. We also summarize the standardized residuals
\begin{equation}
z_{ij}
\equiv
\frac{\theta_{ij}^{\rm true}-\hat{\theta}_{ij}^{\rm HR}}
     {\sigma_{ij}^{\rm HR}}.
\label{eq:npe_sbc_z}
\end{equation}

The rank histograms show an excess near the extremes for $\Omega_m$, $\Omega_b$, $h$, and $n_s$, indicating departures from ideal marginal calibration. The reported standard deviations of $z$ are 1.16, 1.23, 1.29, 1.27, and 0.95 for $(\Omega_m,\Omega_b,h,n_s,\sigma_8)$, respectively. We retain the HR posterior standard deviation as the common unit for the paired shifts. The blue dashed lines in the lower panels mark $\Delta_j=\pm\mathrm{sd}(z_j)$ as an additional reference for the residual dispersion measured in the HR test sample.

The lower panels of Figure~\ref{fig:inference} show the signed shifts for the ten matched pairs. To summarize their magnitudes, we use the
root-mean-square (RMS) normalized shift,
\begin{equation}
D_j^{X}
\equiv
\left[\frac{1}{N_{\rm pair}}
\sum_{i=1}^{N_{\rm pair}}(\Delta_{ij}^{X})^2\right]^{1/2}.
\label{eq:npe_rms}
\end{equation}
For raw SR spectra, the RMS shifts are 2.1, 1.7, and 2.7 in $\Omega_m$, $n_s$, and $\sigma_8$, respectively. For $\Omega_b$ and $h$, the RMS shifts are approximately 0.1. Their small shifts indicate weak sensitivity of the inferred means to the SR--HR differences in this comparison. The $\sigma_8$ shifts are positive in all ten cosmologies, whereas the $\Omega_m$ and $n_s$ shifts vary in sign.

To test how much of the parameter displacement is associated with the model bias measured at the fiducial reference, we apply the same power-spectrum correction as in Section~\ref{subsec:pk}:
\begin{equation}
P_{\rm cal}(k;\boldsymbol{\theta})
=\frac{P_{\rm SR}(k;\boldsymbol{\theta})}{b_{\rm fid}(k)},
\label{eq:npe_calibrated_pk}
\end{equation}
where $b_{\rm fid}(k)$ is the SR-to-HR power ratio defined in Eq.~(\ref{eq:pk_fid_bias}). We then feed the corrected spectra to the same estimator. This reduces the RMS shift in $\sigma_8$ from 2.7 to 0.7, and that in $\Omega_m$ from 2.1 to 1.7, while the RMS shift in $n_s$ remains approximately 1.7. The correction therefore removes a substantial component of the positive $\sigma_8$ shift. The remaining $\Omega_m$ and $n_s$ shifts measure the parameter impact of spectral differences left after the fiducial correction.

The fiducial reference, LH~1128, is itself used to determine $b_{\rm fid}(k)$, so its corrected spectrum matches the HR reference by construction. The other nine pairs test the transfer of this correction to different cosmologies. These results use $k_{\max}=2\,h\,\mathrm{Mpc}^{-1}$, which extends into the nonlinear regime and beyond the Nyquist wavenumber of the particle-grid of the LR, $k_{\rm Nyq}^{\rm LR}\simeq1.6\,h\,\mathrm{Mpc}^{-1}$ (Section~\ref{subsec:pk}). The comparison thus probes parameter sensitivity to the small-scale clustering measured in the SR realizations. A single fiducial power-spectrum calibration brings the $\sigma_8$ estimates substantially closer to the HR reference across the tested cosmologies, while residual shifts in $\Omega_m$ and $n_s$ remain at this choice of $k_{\max}$.

\section{Discussion and Conclusion}
\label{sec:discussion}

The results demonstrate that a single particle-based SR model can reproduce the measured matter clustering and broad changes in halo abundance at cosmologies excluded from the training set. The generator combines the LR phase space fields with explicit conditioning on $(\Omega_m,\Omega_b,h,n_s,\sigma_8)$ and jointly produces particle displacements and velocities. This extends the use of the SR framework across cosmologies, with the same trained model supplying particle realizations for several statistical analyses. The long term objective is simulation-based inference (SBI) using full spatial information learned directly from reconstructed fields. The LR simulation provides the large-scale realization and its non-linear evolution, while the SR model learns to add smaller scale structures to that realization.

Cosmological conditioning is also used in field-level emulators that map linear inputs to non-linear structure \citep{Jamieson2023,Jamieson2024}, and in recent particle SR models \citep{Fremstad2026}. The present work examines a stochastic mapping from evolved LR particle fields, conditioned on all five parameters of the Quijote simulation's LH set, through both clustering and halo population statistics. These comparisons test the combined mapping from the LR fields and cosmological inputs. The different levels of agreement among the statistics provide insight into what the model produces. The visual comparisons and power spectra show that the SR realizations recover the broad morphology and measured two-point statistics of the HR fields. This agreement is obtained without an explicit power spectrum term in the training loss. FoF halo abundances also follow the HR trends, while satellite abundance and occupations exhibit larger and more mass dependent errors. Identifying bound structures tests the re-organization of particles predicted by our SR model within dense environments in addition to the clustering measured by $P(k)$. The satellite results therefore provide information about the generated fields that is complementary to the power spectrum and host-halo abundances.

The suppression of satellite occupation in massive hosts has a similar host mass dependence across the tested cosmologies. A related deficit was found in the SR baseline examined in Paper~IV \citep{Zhang2025}, where the division into particle chunks was investigated as a limitation. This suggests that varying the spatial context of the training chunks, which would require GPUs with more memory, is a concrete next step. Such comparisons would establish whether better resolving structures inside massive hosts improves their satellite populations. Another possible solution is to involve a halo-finding step during training; one could use reinforcement learning to send this non-differentiable feedback to the SR model.

The improvement after fiducial calibration indicates that a substantial part of the SR to HR discrepancy is shared across the tested cosmologies. For each statistic, we derive a scale- or mass-dependent correction from the SR-to-HR ratio at the near-fiducial cosmology and apply it to the other nine test cosmologies. This reduces the power spectrum offset and improves the agreement of FoF halo abundances and satellite occupation in massive hosts. For example, near $k\simeq0.4\,h\,\mathrm{Mpc}^{-1}$, an SR power excess of approximately 2--4\% is reduced to fractional residuals of order 1\% after calibration. Larger residuals remain in the satellite statistics, indicating that the effectiveness of the calibration depends on both the statistic and the scale or mass considered. These corrections are applied to the measured statistics only, leaving the actual generated particle field unchanged. This motivates improvements to the model's understanding of the structure within the actual massive hosts. Such improvements are needed to fulfill our goal of field-level SBI, where cosmological parameters are inferred from information learned directly from the 3D field, without using summary statistics. In this setting, the spatial distribution of particles in small-scale structures can contribute more information beyond the traditional summary statistics tested here.

The NPE comparison connects spectral accuracy to parameter estimation. Although the SR and HR spectra differ by only a few percent over the tested range, replacing an HR spectrum with its SR counterpart can move the inferred parameters by several HR posterior standard deviations. The impact of a spectral residual depends on its scale dependence and on the direction in which the spectrum changes with the cosmological parameters. A small but coherent difference across bins can consequently produce a substantial parameter displacement. After applying calibration to the power spectrum, the reduction of parameter shift shows that a substantial part of this effect is captured by our reference power spectrum correction. The remaining shifts in $\Omega_m$ and $n_s$ show that the accuracy relevant to inference depends on the parameter and on the scales included in the analysis. The motivation for full particle SR is to make the spatial information in those realizations available for inference, including those non-linear structures beyond that described by $P(k)$. Neural compression provides one possible way of combining such features with posterior estimation \citep{Charnock2018,Lemos2024fieldSBI}. An estimator trained on SR fields across parameter space could then be evaluated on independent HR fields.

The stochastic generator also offers a way of producing ensembles of small-scale realizations. Repeated noise draws at fixed LR input would measure the model's conditional scatter, which can be used to construct the distribution and covariance of the generated statistics. The particle outputs make it possible to carry out such checks. All training and evaluation presented in this paper use simulations only at $z = 0$. Combining the present cosmological conditioning with the time-dependent approach of Paper~III would allow joint tests across cosmology and redshift. A different time-conditioning method might be needed as the response of cosmologies and time is inherently different, such as a sinusoidal (Fourier) embedding of the time variable, which is commonly used in diffusion and flow-matching models \citep[e.g.][]{Ho2020}. For galaxy survey applications, the generated matter and halo fields would additionally enter a model for the galaxy population and survey selection, as in forward modeling approaches such as \textsc{SimBIG} \citep{Hahn2023simbig}.

We have extended the AI-assisted particle SR framework to include explicit conditioning on five cosmological parameters. A single model is trained on 100 public Quijote LR--HR pairs at $z = 0$ and evaluated on ten cosmologies excluded from SR training. It maps $512^3$ LR particles to $1024^3$ SR particles in a $1\,h^{-1}\,\mathrm{Gpc}$ box, increasing the particle count by a factor of eight while jointly generating displacements and velocities. The key findings are:

\begin{enumerate}
\item The SR particle projections reproduce the large-scale structures of filaments, dense structures, and voids seen in the matched HR fields across the displayed cosmologies, with visually authentic differences in the detailed structure of individual dense regions.

\item The matter power spectra agree at the few percent level over $0.1<k<3\,h\,\mathrm{Mpc}^{-1}$, and follow the scale-dependent changes in the HR spectra across the test set. The mean of the per-cosmology maximum absolute power errors is 3.6\% over this interval. The comparison extends beyond the LR particle-grid Nyquist wavenumber of $k_{\rm Nyq}^{\rm LR}\simeq1.6\,h\,\mathrm{Mpc}^{-1}$.

\item FoF halo abundances agree to within 15\% in the analyzed mass bins below $10^{14.8}\,h^{-1}M_\odot$. Subhalo abundances have larger mass-dependent errors, and satellite occupation is increasingly suppressed in massive hosts. The fiducial corrections account for a substantial shared component of these statistical errors, with residual satellite discrepancies remaining at the level of tens of percent.

\item At $k_{\max}=2\,h\,\mathrm{Mpc}^{-1}$, the fixed NPE benchmark detects parameter shifts from the raw SR spectra and a clear reduction of the $\sigma_8$ shifts after fiducial calibration. The RMS normalized shift in $\sigma_8$ decreases from 2.7 to 0.7, while residual RMS shifts of approximately 1.7 remain in $\Omega_m$, using the definition in Eq.~(\ref{eq:npe_rms}).
\end{enumerate}

These results show that a cosmology-aware super-resolution model can reproduce matter clustering and halo abundance across the tested cosmologies, and quantify the remaining errors in satellite statistics and parameter inference. Extending the model jointly across cosmology and redshift, improving its substructure predictions, and employing other advanced methods such as diffusion and flow matching are the next steps.

\section*{Data Availability}
The model framework is based on \texttt{map2map} \url{https://github.com/eelregit/map2map}. This PyTorch-based framework is a general-purpose tool for transforming field data. The trained model weights and the pipeline to generate the fields will be available at (\url{https://github.com/xwzhang98/CosmoSR}). Training and test sets data generated in this work will be shared on reasonable request to the corresponding author.

\section*{Acknowledgements}
This work was supported by the Keystone Astronomy and AI (KAAI) program, funded by the Simons Foundation through its Targeted Grants to Institutes program (SFI-MPS-T-Institutes-00021638). We are grateful for discussions with participants of the KAAI program, including Matthew Ho, Barnab\'as P\'oczos, Anshul Kumar, and Yixi Zhang.
We also acknowledge funding from the Block Center for Technology and Society at Carnegie Mellon University.

TDM and RACC acknowledge funding from NASA ATP 80NSSC18K101 and NASA ATP NNX17AK56G. TDM additionally acknowledges funding from NASA ATP 80NSSC20K0519, NSF PHY-2020295, and NASA Theory grant 80NSSC22K072.

The computational results presented have been achieved using Vera at the Pittsburgh Supercomputing Center (PSC).

We also acknowledge the code packages used in this work. The model framework is based on \texttt{map2map} (\url{https://github.com/eelregit/map2map}). Visualization is performed with the open-source code \texttt{gaepsi2} (\url{https://github.com/rainwoodman/gaepsi2}). Data and catalog analysis is performed with the open-source software \texttt{PyTorch} \citep{pytorch} and \texttt{nbodykit} \citep{Hand2018}. ChatGPT (OpenAI) and Claude (Anthropic) were used to help format certain captions and figures and to polish the language; all text generated by these tools was carefully reviewed, edited, and verified for accuracy by the authors, who assume full responsibility for the outcome.

\bibliographystyle{mnras}
\bibliography{refs}

@article{Angulo2021,
  author = {Angulo, Raul E. and Zennaro, Matteo and Contreras, Sergio
            and Aric{\`o}, Giovanni and Pellejero-Iba{\~n}ez, Marcos
            and St{\"u}cker, Jens},
  title = {{The BACCO simulation project: exploiting the full power of
            large-scale structure for cosmology}},
  journal = {MNRAS},
  year = {2021},
  volume = {507},
  number = {4},
  pages = {5869--5881},
  doi = {10.1093/mnras/stab2018},
  archivePrefix = {arXiv},
  eprint = {2004.06245},
  primaryClass = {astro-ph.CO}
}

@ARTICLE{Arjovsky2017,
  author = {Arjovsky, Martin and Chintala, Soumith and Bottou, L{\'e}on},
  title  = {{Wasserstein GAN}},
  journal = {ArXiv e-prints}, year = {2017}, eprint = {1701.07875},
  archivePrefix = {arXiv}, primaryClass = {stat.ML}
}

@ARTICLE{astrid,
       author = {{Bird}, Simeon and {Ni}, Yueying and {Di Matteo}, Tiziana and {Croft}, Rupert and {Feng}, Yu and {Chen}, Nianyi},
        title = "{The ASTRID simulation: galaxy formation and reionization}",
      journal = {\mnras},
         year = 2022,
        month = may,
       volume = {512},
       number = {3},
        pages = {3703-3716},
          doi = {10.1093/mnras/stac648},
archivePrefix = {arXiv},
       eprint = {2111.01160},
 primaryClass = {astro-ph.GA},
       adsurl = {https://ui.adsabs.harvard.edu/abs/2022MNRAS.512.3703B}
}

@ARTICLE{Berger2019,
       author = {{Berger}, Philippe and {Stein}, George},
        title = "{A volumetric deep Convolutional Neural Network for simulation of mock dark matter halo catalogues}",
      journal = {\mnras},
         year = 2019,
        month = jan,
       volume = {482},
       number = {3},
        pages = {2861-2871},
          doi = {10.1093/mnras/sty2949},
archivePrefix = {arXiv},
       eprint = {1805.04537},
 primaryClass = {astro-ph.CO},
       adsurl = {https://ui.adsabs.harvard.edu/abs/2019MNRAS.482.2861B}
}

@ARTICLE{Bernardini2020,
       author = {{Bernardini}, M. and {Mayer}, L. and {Reed}, D. and {Feldmann}, R.},
        title = "{Predicting dark matter halo formation in N-body simulations with deep regression networks}",
      journal = {\mnras},
         year = 2020,
        month = aug,
       volume = {496},
       number = {4},
        pages = {5116-5125},
          doi = {10.1093/mnras/staa1911},
archivePrefix = {arXiv},
       eprint = {1912.04299},
 primaryClass = {astro-ph.CO},
       adsurl = {https://ui.adsabs.harvard.edu/abs/2020MNRAS.496.5116B}
}

@ARTICLE{Bhambra2025,
  author  = {Bhambra, Prabh and Joachimi, Benjamin and Lahav, Ofer and Piras, Davide},
  title   = {{Psi-GAN: a power-spectrum-informed generative adversarial network
             for the emulation of large-scale structure maps across cosmologies
             and redshifts}},
  journal = {MNRAS},
  volume  = {536},
  number  = {3},
  pages   = {3138--3157},
  year    = {2025},
  doi     = {10.1093/mnras/stae2810},
  eprint  = {2410.07349},
  archivePrefix = {arXiv},
  primaryClass  = {astro-ph.CO}
}

@article{Charnock2018,
  author = {Charnock, Tom and Lavaux, Guilhem and Wandelt, Benjamin D.},
  title = {{Automatic physical inference with information maximizing
            neural networks}},
  journal = {Phys. Rev. D},
  year = {2018},
  volume = {97},
  number = {8},
  pages = {083004},
  doi = {10.1103/PhysRevD.97.083004},
  archivePrefix = {arXiv},
  eprint = {1802.03537},
  primaryClass = {astro-ph.IM}
}

@ARTICLE{Crocce2006,
       author = {{Crocce}, Mart{\'\i}n and {Pueblas}, Sebasti{\'a}n and {Scoccimarro}, Rom{\'a}n},
        title = "{Transients from initial conditions in cosmological simulations}",
      journal = {\mnras},
         year = 2006,
        month = nov,
       volume = {373},
       number = {1},
        pages = {369-381},
          doi = {10.1111/j.1365-2966.2006.11040.x},
archivePrefix = {arXiv},
       eprint = {astro-ph/0606505},
 primaryClass = {astro-ph},
       adsurl = {https://ui.adsabs.harvard.edu/abs/2006MNRAS.373..369C}
}

@ARTICLE{Davis85,
       author = {{Davis}, M. and {Efstathiou}, G. and {Frenk}, C.~S. and
         {White}, S.~D.~M.},
        title = "{The evolution of large-scale structure in a universe dominated by cold dark matter}",
      journal = {\apj},
         year = "1985",
        month = "May",
       volume = {292},
        pages = {371-394},
          doi = {10.1086/163168},
       adsurl = {https://ui.adsabs.harvard.edu/abs/1985ApJ...292..371D}
}

@ARTICLE{DESI2016,
  author = {{DESI Collaboration} and others},
  title  = {{The DESI Experiment Part I: Science, Targeting, and Survey Design}},
  journal = {ArXiv e-prints}, year = {2016}, eprint = {1611.00036},
  archivePrefix = {arXiv}, primaryClass = {astro-ph.IM}
}

@ARTICLE{Dvorkin2022,
  author  = {Dvorkin, Cora and others},
  title   = {{Machine Learning and Cosmology}},
  journal = {ArXiv e-prints},
  year    = {2022},
  eprint  = {2203.08056},
  archivePrefix = {arXiv},
  primaryClass  = {astro-ph.CO}
}

@ARTICLE{EuclidFlagship2025,
  author  = {{Euclid Collaboration} and others},
  title   = {{Euclid. V. The Flagship galaxy mock catalogue:
             A comprehensive simulation for the Euclid mission}},
  journal = {A\&A},
  year    = {2025},
  volume  = {697},
  pages   = {A5},
  doi     = {10.1051/0004-6361/202450853},
  eprint  = {2405.13495},
  archivePrefix = {arXiv},
  primaryClass  = {astro-ph.CO}
}

@ARTICLE{fastpm,
       author = {{Feng}, Yu and {Chu}, Man-Yat and {Seljak}, Uro{\v{s}} and {McDonald}, Patrick},
        title = "{FASTPM: a new scheme for fast simulations of dark matter and haloes}",
      journal = {\mnras},
         year = 2016,
        month = dec,
       volume = {463},
       number = {3},
        pages = {2273-2286},
          doi = {10.1093/mnras/stw2123},
archivePrefix = {arXiv},
       eprint = {1603.00476},
 primaryClass = {astro-ph.CO},
       adsurl = {https://ui.adsabs.harvard.edu/abs/2016MNRAS.463.2273F}
}

@ARTICLE{Fremstad2026,
  author  = {Fremstad, Dennis and Adamek, Julian and Mota, David F.},
  title   = {{Separate Universe Super-Resolution Emulator}},
  journal = {ArXiv e-prints},
  year    = {2026},
  eprint  = {2605.09004},
  archivePrefix = {arXiv},
  primaryClass  = {astro-ph.CO}
}

@article{Frontiere2022,
  author = {Frontiere, Nicholas and Heitmann, Katrin and Rangel, Esteban
            and Larsen, Patricia and Pope, Adrian and Sultan, Imran
            and Uram, Thomas and Habib, Salman and Rizzi, Silvio
            and Insley, Joe},
  title = {{Farpoint: A High-resolution Cosmology Simulation
            at the Gigaparsec Scale}},
  journal = {ApJS},
  year = {2022},
  volume = {259},
  number = {1},
  pages = {15},
  doi = {10.3847/1538-4365/ac43b9},
  archivePrefix = {arXiv},
  eprint = {2109.01956},
  primaryClass = {astro-ph.CO}
}

@ARTICLE{Goodfellow2014,
  author = {Goodfellow, Ian J. and Pouget-Abadie, Jean and Mirza, Mehdi and Xu, Bing
            and Warde-Farley, David and Ozair, Sherjil and Courville, Aaron and Bengio, Yoshua},
  title  = {{Generative Adversarial Networks}},
  journal = {ArXiv e-prints}, year = {2014}, eprint = {1406.2661},
  archivePrefix = {arXiv}, primaryClass = {stat.ML}
}

@INPROCEEDINGS{Gulrajani2017,
  author = {Gulrajani, Ishaan and Ahmed, Faruk and Arjovsky, Martin and
            Dumoulin, Vincent and Courville, Aaron C.},
  title  = {{Improved Training of Wasserstein GANs}},
  booktitle = {Advances in Neural Information Processing Systems (NeurIPS)},
  year   = {2017}, eprint = {1704.00028}, archivePrefix = {arXiv},
  primaryClass = {cs.LG}
}

@article{Hahn2023simbig,
  author = {Hahn, ChangHoon and Eickenberg, Michael and Ho, Shirley
            and Hou, Jiamin and Lemos, Pablo and Massara, Elena and Modi, Chirag
            and {Moradinezhad Dizgah}, Azadeh
            and {R{\'e}galdo-Saint Blancard}, Bruno and Abidi, Muntazir M.},
  title = {{A forward modeling approach to analyzing galaxy clustering
            with SimBIG}},
  journal = {Proc. Natl. Acad. Sci.},
  year = {2023},
  volume = {120},
  number = {42},
  pages = {e2218810120},
  doi = {10.1073/pnas.2218810120},
  archivePrefix = {arXiv},
  eprint = {2211.00723},
  primaryClass = {astro-ph.CO}
}

@ARTICLE{Hand2018,
       author = {{Hand}, Nick and {Feng}, Yu and {Beutler}, Florian and {Li}, Yin and {Modi}, Chirag and {Seljak}, Uro{\v{s}} and {Slepian}, Zachary},
        title = "{nbodykit: An Open-source, Massively Parallel Toolkit for Large-scale Structure}",
      journal = {\aj},
         year = 2018,
        month = oct,
       volume = {156},
       number = {4},
          eid = {160},
        pages = {160},
          doi = {10.3847/1538-3881/aadae0},
archivePrefix = {arXiv},
       eprint = {1712.05834},
 primaryClass = {astro-ph.IM},
       adsurl = {https://ui.adsabs.harvard.edu/abs/2018AJ....156..160H}
}

@article{He2019,
  author = {He, Siyu and Li, Yin and Feng, Yu and Ho, Shirley
            and Ravanbakhsh, Siamak and Chen, Wei and P{\'o}czos, Barnab{\'a}s},
  title = {{Learning to predict the cosmological structure formation}},
  journal = {Proc. Natl. Acad. Sci.},
  year = {2019},
  volume = {116},
  number = {28},
  pages = {13825--13832},
  doi = {10.1073/pnas.1821458116},
  archivePrefix = {arXiv},
  eprint = {1811.06533},
  primaryClass = {astro-ph.CO}
}

@ARTICLE{Ho2020,
  author = {Ho, Jonathan and Jain, Ajay and Abbeel, Pieter},
  title  = {{Denoising Diffusion Probabilistic Models}},
  journal = {ArXiv e-prints}, year = {2020}, eprint = {2006.11239},
  archivePrefix = {arXiv}, primaryClass = {cs.LG}
}

@ARTICLE{Hopkins2014,
  author  = {Hopkins, Philip F. and Kere{\v s}, Du{\v s}an and O{\~n}orbe, Jos{\'e} and
             Faucher-Gigu{\`e}re, Claude-Andr{\'e} and Quataert, Eliot and
             Murray, Norman and Bullock, James S.},
  title   = {{Galaxies on FIRE (Feedback In Realistic Environments): stellar
             feedback explains cosmologically inefficient star formation}},
  journal = {MNRAS},
  volume  = {445},
  number  = {1},
  pages   = {581--603},
  year    = {2014},
  doi     = {10.1093/mnras/stu1738},
  eprint  = {1311.2073},
  archivePrefix = {arXiv},
  primaryClass  = {astro-ph.CO}
}

@ARTICLE{HuertasCompany2023,
  author = {Huertas-Company, Marc and Lanusse, Fran{\c c}ois},
  title  = {{The Dawes Review 10: The impact of deep learning for the analysis
            of galaxy surveys}},
  journal = {PASA}, volume = {40}, pages = {e001}, year = {2023},
  doi = {10.1017/pasa.2022.55}, eprint = {2210.01813},
  archivePrefix = {arXiv}, primaryClass = {astro-ph.GA}
}

@ARTICLE{hydroEmu,
       author = {{Hafezianzadeh}, Fatemeh and {Zhang}, Xiaowen and {Ni}, Yueying and {Croft}, Rupert A.~C. and {DiMatteo}, Tiziana and {Qezlou}, Mahdi and {Bird}, Simeon},
        title = "{An AI Superresolution Field Emulator for Cosmological Hydrodynamics: The Ly{\ensuremath{\alpha}} Forest}",
      journal = {\apj},
         year = 2025,
        month = dec,
       volume = {995},
       number = {1},
          eid = {42},
        pages = {42},
          doi = {10.3847/1538-4357/ae1474},
archivePrefix = {arXiv},
       eprint = {2507.16189},
 primaryClass = {astro-ph.CO},
       adsurl = {https://ui.adsabs.harvard.edu/abs/2025ApJ...995...42H}
}

@ARTICLE{Ivezic2019,
  author  = {Ivezi{\'c}, {\v Z}eljko and Kahn, Steven M. and Tyson, J. Anthony and others},
  title   = {{LSST: From Science Drivers to Reference Design and Anticipated
             Data Products}},
  journal = {ApJ},
  volume  = {873},
  number  = {2},
  pages   = {111},
  year    = {2019},
  doi     = {10.3847/1538-4357/ab042c},
  eprint  = {0805.2366},
  archivePrefix = {arXiv},
  primaryClass  = {astro-ph}
}

@ARTICLE{Jamieson2023,
  author  = {Jamieson, Drew and Li, Yin and {Alves de Oliveira}, Renan and
             Villaescusa-Navarro, Francisco and Ho, Shirley and Spergel, David N.},
  title   = {{Field-level Neural Network Emulator for Cosmological N-body Simulations}},
  journal = {ApJ},
  volume  = {952},
  number  = {2},
  pages   = {145},
  year    = {2023},
  doi     = {10.3847/1538-4357/acdb6c},
  eprint  = {2206.04594},
  archivePrefix = {arXiv},
  primaryClass  = {astro-ph.CO}
}

@ARTICLE{Jamieson2024,
  author  = {Jamieson, Drew and Li, Yin and Villaescusa-Navarro, Francisco
             and Ho, Shirley and Spergel, David N.},
  title   = {{Field-level emulation of cosmic structure formation with
             cosmology and redshift dependence}},
  journal = {JCAP},
  year    = {2025},
  volume  = {2025},
  number  = {03},
  pages   = {072},
  doi     = {10.1088/1475-7516/2025/03/072},
  eprint  = {2408.07699},
  archivePrefix = {arXiv},
  primaryClass  = {astro-ph.CO}
}

@ARTICLE{Karras2019,
  author = {Karras, Tero and Laine, Samuli and Aittala, Miika and Hellsten, Janne
            and Lehtinen, Jaakko and Aila, Timo},
  title  = {{Analyzing and Improving the Image Quality of StyleGAN}},
  journal = {ArXiv e-prints}, year = {2019}, eprint = {1912.04958},
  archivePrefix = {arXiv}, primaryClass = {cs.CV}
}

@INPROCEEDINGS{KingmaBa2015,
  author = {Kingma, Diederik P. and Ba, Jimmy},
  title  = {{Adam: A Method for Stochastic Optimization}},
  booktitle = {International Conference on Learning Representations (ICLR)},
  year   = {2015}, eprint = {1412.6980}, archivePrefix = {arXiv},
  primaryClass = {cs.LG}
}

@ARTICLE{KodiRamanah2020,
       author = {{Kodi Ramanah}, Doogesh and {Charnock}, Tom and {Villaescusa-Navarro}, Francisco and {Wandelt}, Benjamin D.},
        title = "{Super-resolution emulator of cosmological simulations using deep physical models}",
      journal = {\mnras},
         year = 2020,
        month = jul,
       volume = {495},
       number = {4},
        pages = {4227-4236},
          doi = {10.1093/mnras/staa1428},
archivePrefix = {arXiv},
       eprint = {2001.05519},
 primaryClass = {astro-ph.CO},
       adsurl = {https://ui.adsabs.harvard.edu/abs/2020MNRAS.495.4227K}
}

@ARTICLE{Laureijs2011,
  author = {Laureijs, R. and others},
  title  = {{Euclid Definition Study Report}},
  journal = {ArXiv e-prints}, year = {2011}, eprint = {1110.3193},
  archivePrefix = {arXiv}, primaryClass = {astro-ph.CO}
}

@ARTICLE{Ledig2016,
       author = {{Ledig}, Christian and {Theis}, Lucas and {Huszar}, Ferenc and {Caballero}, Jose and {Cunningham}, Andrew and {Acosta}, Alejandro and {Aitken}, Andrew and {Tejani}, Alykhan and {Totz}, Johannes and {Wang}, Zehan and {Shi}, Wenzhe},
        title = "{Photo-Realistic Single Image Super-Resolution Using a Generative Adversarial Network}",
      journal = {ArXiv e-prints},
         year = 2016,
        month = sep,
archivePrefix = {arXiv},
       eprint = {1609.04802},
 primaryClass = {cs.CV}
}

@ARTICLE{Lemos2024fieldSBI,
  author  = {Lemos, Pablo and Parker, Liam and Hahn, ChangHoon
             and Ho, Shirley and Eickenberg, Michael and Hou, Jiamin
             and Massara, Elena and Modi, Chirag
             and {Moradinezhad Dizgah}, Azadeh
             and {R{\'e}galdo-Saint Blancard}, Bruno and Spergel, David},
  title   = {{Field-level simulation-based inference of galaxy clustering
             with convolutional neural networks}},
  journal = {Phys. Rev. D},
  year    = {2024},
  volume  = {109},
  number  = {8},
  pages   = {083536},
  doi     = {10.1103/PhysRevD.109.083536},
  eprint  = {2310.15256},
  archivePrefix = {arXiv},
  primaryClass  = {astro-ph.CO}
}

@article{Lewis2000,
  author = {Lewis, Antony and Challinor, Anthony and Lasenby, Anthony},
  title = {{Efficient Computation of CMB Anisotropies in Closed FRW Models}},
  journal = {ApJ},
  year = {2000},
  volume = {538},
  pages = {473--476},
  doi = {10.1086/309179},
  archivePrefix = {arXiv},
  eprint = {astro-ph/9911177},
  primaryClass = {astro-ph}
}

@article{Li2021,
  author = {Li, Yin and Ni, Yueying and Croft, Rupert A. C.
            and {Di Matteo}, Tiziana and Bird, Simeon and Feng, Yu},
  title = {{AI-assisted superresolution cosmological simulations}},
  journal = {Proc. Natl. Acad. Sci.},
  year = {2021},
  volume = {118},
  number = {19},
  pages = {e2022038118},
  doi = {10.1073/pnas.2022038118},
  archivePrefix = {arXiv},
  eprint = {2010.06608},
  primaryClass = {astro-ph.CO}
}

@ARTICLE{Maksimova2021,
  author  = {Maksimova, Nina A. and Garrison, Lehman H. and Eisenstein, Daniel J.
             and Hadzhiyska, Boryana and Bose, Sownak and Satterthwaite, Thomas P.},
  title   = {{ABACUSSUMMIT: a massive set of high-accuracy, high-resolution
             N-body simulations}},
  journal = {MNRAS},
  volume  = {508},
  number  = {3},
  pages   = {4017--4037},
  year    = {2021},
  doi     = {10.1093/mnras/stab2484},
  eprint  = {2110.11398},
  archivePrefix = {arXiv},
  primaryClass  = {astro-ph.CO}
}

@ARTICLE{Mirza2014,
  author = {Mirza, Mehdi and Osindero, Simon},
  title  = {{Conditional Generative Adversarial Nets}},
  journal = {ArXiv e-prints}, year = {2014}, eprint = {1411.1784},
  archivePrefix = {arXiv}, primaryClass = {cs.LG}
}

@ARTICLE{Ni2021,
  author  = {Ni, Yueying and Li, Yin and Lachance, Patrick and Croft, Rupert A. C.
             and {Di Matteo}, Tiziana and Bird, Simeon and Feng, Yu},
  title   = {{AI-assisted superresolution cosmological simulations II. Halo
             substructures, velocities, and higher order statistics}},
  journal = {MNRAS},
  volume  = {507},
  number  = {1},
  pages   = {1021--1033},
  year    = {2021},
  doi     = {10.1093/mnras/stab2113},
  eprint  = {2105.01016},
  archivePrefix = {arXiv},
  primaryClass  = {astro-ph.CO}
}

@article{Nishimichi2019,
  author = {Nishimichi, Takahiro and Takada, Masahiro and Takahashi, Ryuichi
            and Osato, Ken and Shirasaki, Masato and Oogi, Taira
            and Miyatake, Hironao and Oguri, Masamune and Murata, Ryoma
            and Kobayashi, Yosuke and Yoshida, Naoki},
  title = {{Dark Quest. I. Fast and Accurate Emulation of Halo Clustering
            Statistics and Its Application to Galaxy Clustering}},
  journal = {ApJ},
  year = {2019},
  volume = {884},
  number = {1},
  pages = {29},
  doi = {10.3847/1538-4357/ab3719},
  archivePrefix = {arXiv},
  eprint = {1811.09504},
  primaryClass = {astro-ph.CO}
}

@inproceedings{Papamakarios2016,
  author = {Papamakarios, George and Murray, Iain},
  title = {{Fast $\epsilon$-free Inference of Simulation Models with
            Bayesian Conditional Density Estimation}},
  booktitle = {Advances in Neural Information Processing Systems},
  volume = {29},
  year = {2016},
  url = {https://proceedings.neurips.cc/paper/2016/hash/6aca97005c68f1206823815f66102863-Abstract.html},
  archivePrefix = {arXiv},
  eprint = {1605.06376},
  primaryClass = {stat.ML}
}

@ARTICLE{Perraudin2019,
       author = {{Perraudin}, Nathana{\"e}l and {Srivastava}, Ankit and {Lucchi}, Aurelien and {Kacprzak}, Tomasz and {Hofmann}, Thomas and {R{\'e}fr{\'e}gier}, Alexandre},
        title = "{Cosmological N-body simulations: a challenge for scalable generative models}",
      journal = {Computational Astrophysics and Cosmology},
         year = 2019,
        month = dec,
       volume = {6},
       number = {1},
          eid = {5},
        pages = {5},
          doi = {10.1186/s40668-019-0032-1},
archivePrefix = {arXiv},
       eprint = {1908.05519},
 primaryClass = {physics.comp-ph},
       adsurl = {https://ui.adsabs.harvard.edu/abs/2019ComAC...6....5P}
}

@ARTICLE{Perraudin2020,
       author = {{Perraudin}, Nathana{\"e}l and {Marcon}, Sandro and {Lucchi}, Aurelien and {Kacprzak}, Tomasz},
        title = "{Emulation of cosmological mass maps with conditional generative adversarial networks}",
      journal = {ArXiv e-prints},
         year = 2020,
        month = apr,
archivePrefix = {arXiv},
       eprint = {2004.08139},
 primaryClass = {astro-ph.CO}
}

@ARTICLE{pytorch,
       author = {{Paszke}, Adam and {Gross}, Sam and {Massa}, Francisco and {Lerer}, Adam and {Bradbury}, James and {Chanan}, Gregory and {Killeen}, Trevor and {Lin}, Zeming and {Gimelshein}, Natalia and {Antiga}, Luca and {Desmaison}, Alban and {K{\"o}pf}, Andreas and {Yang}, Edward and {DeVito}, Zach and {Raison}, Martin and {Tejani}, Alykhan and {Chilamkurthy}, Sasank and {Steiner}, Benoit and {Fang}, Lu and {Bai}, Junjie and {Chintala}, Soumith},
        title = "{PyTorch: An Imperative Style, High-Performance Deep Learning Library}",
      journal = {ArXiv e-prints},
         year = 2019,
        month = dec,
archivePrefix = {arXiv},
       eprint = {1912.01703},
 primaryClass = {cs.LG}
}

@ARTICLE{Rodriguez2018,
  author  = {Rodr{\'i}guez, Andres C. and Kacprzak, Tomasz and Lucchi, Aurelien and
             Amara, Adam and Sgier, Rapha{\"e}l and Fluri, Janis and
             Hofmann, Thomas and R{\'e}fr{\'e}gier, Alexandre},
  title   = {{Fast cosmic web simulations with generative adversarial networks}},
  journal = {Comput. Astrophys. Cosmol.},
  volume  = {5},
  number  = {1},
  pages   = {4},
  year    = {2018},
  doi     = {10.1186/s40668-018-0026-4},
  eprint  = {1801.09070},
  archivePrefix = {arXiv},
  primaryClass  = {astro-ph.CO}
}

@article{Rouhiainen2024,
  author = {Rouhiainen, Adam and M{\"u}nchmeyer, Moritz and Shiu, Gary
            and Gira, Michael and Lee, Kangwook},
  title = {{Superresolution emulation of large cosmological fields with
            a 3D conditional diffusion model}},
  journal = {Phys. Rev. D},
  year = {2024},
  volume = {109},
  number = {12},
  pages = {123536},
  doi = {10.1103/PhysRevD.109.123536},
  archivePrefix = {arXiv},
  eprint = {2311.05217},
  primaryClass = {astro-ph.CO}
}

@article{Schanz2024,
  author = {Schanz, Andreas and List, Florian and Hahn, Oliver},
  title = {{Stochastic Super-resolution of Cosmological Simulations with
            Denoising Diffusion Models}},
  journal = {The Open Journal of Astrophysics},
  year = {2024},
  volume = {7},
  doi = {10.33232/001c.125902},
  archivePrefix = {arXiv},
  eprint = {2310.06929},
  primaryClass = {astro-ph.CO}
}

@ARTICLE{Sipp2023,
  author  = {Sipp, Meris and LaChance, Patrick and Croft, Rupert and
             Ni, Yueying and {Di Matteo}, Tiziana},
  title   = {{Towards super-resolution simulations of the fuzzy dark matter
             cosmological model}},
  journal = {MNRAS},
  volume  = {525},
  number  = {2},
  pages   = {1807--1813},
  year    = {2023},
  doi     = {10.1093/mnras/stad2341},
  eprint  = {2210.12907},
  archivePrefix = {arXiv},
  primaryClass  = {astro-ph.CO}
}

@ARTICLE{SohlDickstein2015,
  author = {Sohl-Dickstein, Jascha and Weiss, Eric A. and Maheswaranathan, Niru and Ganguli, Surya},
  title  = {{Deep Unsupervised Learning using Nonequilibrium Thermodynamics}},
  journal = {ArXiv e-prints}, year = {2015}, eprint = {1503.03585},
  archivePrefix = {arXiv}, primaryClass = {cs.LG}
}

@ARTICLE{Springel2001subfind,
  author  = {Springel, Volker and White, Simon D. M. and Tormen, Giuseppe and Kauffmann, Guinevere},
  title   = {{Populating a cluster of galaxies -- I. Results at z=0}},
  journal = {MNRAS},
  volume  = {328},
  number  = {3},
  pages   = {726--750},
  year    = {2001},
  doi     = {10.1046/j.1365-8711.2001.04912.x},
  eprint  = {astro-ph/0012055},
  archivePrefix = {arXiv}
}

@ARTICLE{Springel2005,
  author  = {Springel, Volker},
  title   = {{The cosmological simulation code GADGET-2}},
  journal = {MNRAS},
  volume  = {364},
  number  = {4},
  pages   = {1105--1134},
  year    = {2005},
  doi     = {10.1111/j.1365-2966.2005.09655.x},
  eprint  = {astro-ph/0505010},
  archivePrefix = {arXiv}
}

@article{Talts2018,
  author = {Talts, Sean and Betancourt, Michael and Simpson, Daniel
            and Vehtari, Aki and Gelman, Andrew},
  title = {{Validating Bayesian Inference Algorithms with
            Simulation-Based Calibration}},
  journal = {ArXiv e-prints},
  year = {2018},
  archivePrefix = {arXiv},
  eprint = {1804.06788},
  primaryClass = {stat.ME},
  url = {https://arxiv.org/abs/1804.06788}
}

@article{Tassev2013,
  author = {Tassev, Svetlin and Zaldarriaga, Matias and Eisenstein, Daniel J.},
  title = {{Solving large scale structure in ten easy steps with COLA}},
  journal = {JCAP},
  year = {2013},
  volume = {2013},
  number = {06},
  pages = {036},
  doi = {10.1088/1475-7516/2013/06/036},
  archivePrefix = {arXiv},
  eprint = {1301.0322},
  primaryClass = {astro-ph.CO}
}

@ARTICLE{VillaescusaNavarro2020,
  author  = {Villaescusa-Navarro, Francisco and others},
  title   = {{The Quijote Simulations}},
  journal = {ApJS},
  volume  = {250},
  number  = {1},
  pages   = {2},
  year    = {2020},
  doi     = {10.3847/1538-4365/ab9d82},
  eprint  = {1909.05273},
  archivePrefix = {arXiv},
  primaryClass  = {astro-ph.CO}
}

@ARTICLE{VillaescusaNavarro2020camels,
  author = {Villaescusa-Navarro, Francisco and others},
  title  = {{The CAMELS Project: Cosmology and Astrophysics with Machine-learning
            Simulations}},
  journal = {ApJ}, volume = {915}, number = {1}, pages = {71}, year = {2021},
  doi = {10.3847/1538-4357/abf7ba}, eprint = {2010.00619},
  archivePrefix = {arXiv}, primaryClass = {astro-ph.CO}
}

@ARTICLE{Wang2019,
  author = {Wang, Zhihao and Chen, Jian and Hoi, Steven C. H.},
  title  = {{Deep Learning for Image Super-resolution: A Survey}},
  journal = {ArXiv e-prints}, year = {2019}, eprint = {1902.06068},
  archivePrefix = {arXiv}, primaryClass = {cs.CV}
}

@ARTICLE{Zhang2023,
  author  = {Zhang, Xiaowen and Lachance, Patrick and Ni, Yueying and Li, Yin and
             Croft, Rupert A. C. and {Di Matteo}, Tiziana and Bird, Simeon and Feng, Yu},
  title   = {{AI-assisted super-resolution cosmological simulations III: time evolution}},
  journal = {MNRAS},
  volume  = {528},
  number  = {1},
  pages   = {281--293},
  year    = {2024},
  doi     = {10.1093/mnras/stad3940},
  eprint  = {2305.12222},
  archivePrefix = {arXiv},
  primaryClass  = {astro-ph.CO}
}

@ARTICLE{Zhang2025,
  author  = {Zhang, Xiaowen and Lachance, Patrick and Dasgupta, Ankita and
             Croft, Rupert A. C. and {Di Matteo}, Tiziana and Ni, Yueying and
             Bird, Simeon and Li, Yin},
  title   = {{AI-assisted super-resolution cosmological simulations IV: An
             emulator for deterministic realizations}},
  journal = {The Open Journal of Astrophysics},
  volume  = {8},
  year    = {2025},
  doi     = {10.33232/001c.129471},
  eprint  = {2408.09051},
  archivePrefix = {arXiv},
  primaryClass  = {astro-ph.CO}
}

@article{Zhou2026Astrid,
  author = {Zhou, Yihao and {Di Matteo}, Tiziana and Bird, Simeon
            and Croft, Rupert and Ni, Yueying and Yang, Yanhui
            and Chen, Nianyi and Lachance, Patrick and Zhang, Xiaowen
            and Hafezianzadeh, Fatemeh},
  title = {{The ASTRID Simulation at z=0: From Massive Black Holes
            to Large-scale Structure}},
  journal = {ApJ},
  year = {2026},
  volume = {999},
  number = {1},
  pages = {41},
  doi = {10.3847/1538-4357/ae3c08},
  archivePrefix = {arXiv},
  eprint = {2510.13976},
  primaryClass = {astro-ph.GA}
}
\end{document}